\documentclass[twocolumn,prl,twocolumn,superscriptaddress]{revtex4-1}
\usepackage{amsmath,amssymb,mathrsfs}
\usepackage{natbib}
\usepackage{subfigure}
\usepackage{tabularx}
\usepackage{epsfig}
\usepackage{longtable}
\usepackage{amsfonts}
\usepackage{rotating}
\usepackage{subfigure}
\usepackage{amsmath}
\usepackage{comment}
\usepackage{bbold}

\def\be{\begin{equation}}
\def\ee{\end{equation}}
\def\bea{\begin{eqnarray}}
\def\eea{\end{eqnarray}}
\def\nn{\nonumber}

\usepackage{wordlike}

\usepackage[unicode=true,bookmarks=true,bookmarksnumbered=false,bookmarksopen=false,breaklinks=false,pdfborder={0 0 1},backref=false,colorlinks=true]{hyperref}

\hypersetup{linkcolor=magenta,urlcolor=blue,citecolor=blue,pdfstartview={FitH},hyperfootnotes=false,unicode=true}

\begin{document}

\title{Evidence of Potts-Nematic Superfluidity in a Hexagonal $sp^2$ Optical Lattice}

\author{Shengjie Jin}
\affiliation{State Key Laboratory of Advanced Optical Communication System and Network, Department of Electronics, Peking University, Beijing 100871, China}
\author{Wenjun Zhang}
\affiliation{State Key Laboratory of Advanced Optical Communication System and Network, Department of Electronics, Peking University, Beijing 100871, China}
\author{Xinxin Guo}
\affiliation{State Key Laboratory of Advanced Optical Communication System and Network, Department of Electronics, Peking University, Beijing 100871, China}
\author{Xuzong Chen}
\affiliation{State Key Laboratory of Advanced Optical Communication System and Network, Department of Electronics, Peking University, Beijing 100871, China}
\author{Xiaoji Zhou}
\email{xjzhou@pku.edu.cn}
\affiliation{State Key Laboratory of Advanced Optical Communication System and Network, Department of Electronics, Peking University, Beijing 100871, China}
\affiliation{Collaborative Innovation Center of Extreme Optics, Shanxi University, Taiyuan, Shanxi 030006, China}
\author{Xiaopeng Li}
\email{xiaopeng\_li@fudan.edu.cn}
\affiliation{State Key Laboratory of Surface Physics, Institute of Nanoelectronics and Quantum Computing, Department of Physics, Fudan University, Shanghai 200438, China}
\affiliation{Shanghai Qizhi Institute, AI Tower, Xuhui District, Shanghai 200232, China}
\date{\today}

\begin{abstract}
As in between liquid and crystal phases lies a nematic  liquid crystal, which breaks rotation with preservation of translation symmetry, there is a nematic superfluid phase bridging a superfluid and a supersolid. The nematic order also emerges in interacting electrons and has been found to largely intertwine with multi-orbital correlation in high-temperature superconductivity, where Ising nematicity arises from a four-fold rotation symmetry $C_4$  broken down to $C_2$. 
 Here we report an observation of a three-state ($\mathbb{Z}_3$) quantum nematic order, dubbed ``Potts-nematicity", 
 in a system of cold atoms loaded in an excited band of a hexagonal optical lattice described by an $sp^2$-orbital hybridized model. This Potts-nematic quantum state spontaneously breaks a three-fold rotation symmetry of the lattice, qualitatively distinct from the Ising nematicity. Our field theory analysis shows that the Potts-nematic order is stabilized by intricate renormalization effects  enabled by strong  inter-orbital mixing present in the hexagonal lattice. This discovery paves a way to investigate quantum  vestigial orders in multi-orbital atomic superfluids. 
\end{abstract}

\maketitle

In electronic materials, the existence of nematic order~\cite{2000_Chaikin_Book} has been established in high temperature superconductors such as cuprates~\cite{2010_Fradkin_CMP} and iron-based superconductors~\cite{2010_Fisher_Science,2014_Schmalian_NatPhys,2014_Schmalian_NatPhys,2016_Si_NRM,2019intertwined_Schmalian_ARCMP}. The quantum liquid crystal phase is  of great importance  to the fundamental understanding of high temperature superconductivity.
The investigation of intertwined vestigial orders in multi-orbital superconductivity that incorporates nematicity  has been attracting much attention~\cite{2019intertwined_Schmalian_ARCMP} in recent years. In these superconducting materials, an Ising nematic order is most predominantly  observed, where the nematic orientation has only two choices. In such systems, what drives the nematic order has ambiguity for it is difficult to separate the electron correlation effects from material structural transitions~\cite{2014_Schmalian_NatPhys}.

The system of ultracold neutral atoms confined in optical lattices has a large degree of controllability. The backaction from atoms to the confining laser potential is typically negligible, making the structural transition avoidable. As an effort to build an optical lattice emulator for multi-orbital physics~\cite{2011_Lewenstein_NatPhys,2016_Li_RPP},  excited band condensation of cold atoms  has been achieved in one~\cite{2013_Chin_NatPhys,2018_Zhou_PRL} and two-dimensional lattices~\cite{2011_Hemmerich_NatPhys,2015_Hemmerich_PRL,2011_Sengstock_Science,2012_Sengstock_NatPhys}.  A crucial difference of such condensates from the ground-state condensate is the physics is generically described by a multi-component order parameter that respects crystalline symmetries~\cite{2011_Lewenstein_NatPhys,2016_Li_RPP}, distinctive from single-component~\cite{2008_Pethick_BEC} or multi-component spinor condensates~\cite{2013_Samper-Kurn_RMP}. At the level of effective field theory description, this atomic system shares similarity as multi-orbital iron-based superconductors and enjoys more controllability.   Interaction driven orbital orders such as chiral symmetry breaking~\cite{2011_Hemmerich_NatPhys,2015_Hemmerich_PRL,2012_Sengstock_NatPhys,2016_Li_RPP}, and dynamical phase-sliding~\cite{2018_Zhou_PRL}  
have been reported in multi-orbital settings of cold atoms. 
But many-body correlation effects beyond mean field theory have not been observed so far in such experimental systems.

\begin{figure*}[htp]
\centerline{\includegraphics[angle=0,width=0.9\textwidth]{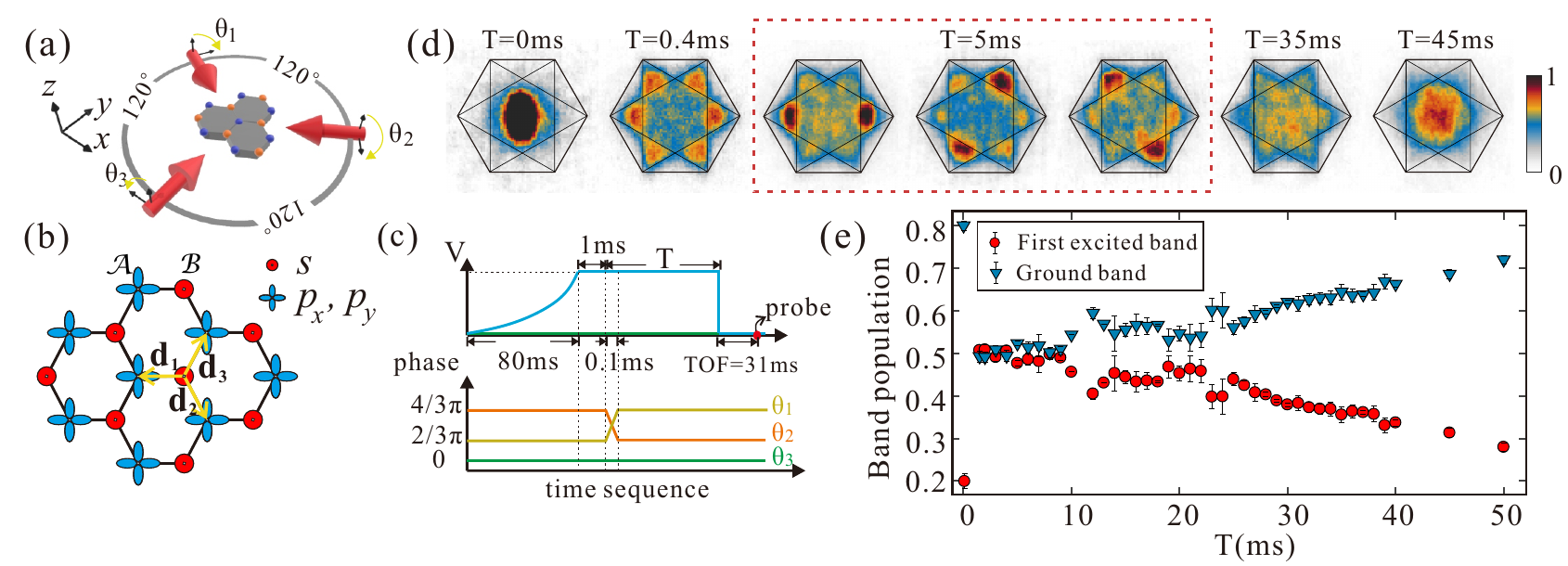}}
\caption{
{\bf Experimental protocol of loading atoms into the excited $sp^2$-band of the hexagonal optical lattice.}
{\bf a}, illustrates the arrangement of the laser beams forming the hexagonal lattice. There are three laser beams in the x-y plane with laser-wavelength
$\lambda = 1064$nm forming an optical hexagonal lattice~\cite{supplement}. The three angles $\theta_{1,2,3}$ represent the relative phases of the elliptical polarization of the light.
{\bf b}, the geometry of the hexagonal lattice, having two sets of sublattices labeled by  ${\cal A}$ and ${\cal B}$. 
The lattice is formed taking $\{\theta_1,\theta_2,\theta_3\}=\{4\pi/3, 2\pi/3,0\}$ or $\{2\pi/3, 4\pi/3,0\}$.
{\bf c}, The time sequence implemented in the experiment to load atoms from the lowest to the second band. 
{\bf d}, the band-mapping TOF images with $T=0, 0.4, 5, 35, 45\rm{ms}$. When $T=5\rm{ms}$, three different band-mapping TOF corresponding three nematic orientations are shown in the red dashed box.
{\bf e}, The measured time-evolution of the atomic population in the ground and first-excited bands normalized by their sum.
Here we average over five experimental runs for each data point, with the error bar denoting the statistical error.
}
\label{fig:f1}
\end{figure*}

Here we report observation of a Potts-nematic quantum state  in a system of  cold atoms loaded into the second band of a hexagonal optical lattice. 
The emergence of this novel phase is not captured by a simple mean field theory.
We first prepare an atomic  Bose-Einstein condensate  (BEC) in the ground band which respects all symmetries of the lattice, 
and then project the atomic sample onto the band-maximum of the second band using a lattice quench (see Fig.~\ref{fig:f1}). 
The phase coherence in the state will immediately disappear and then reemerge within a few milliseconds.
During this process of phase-coherence reformation, the quantum state spontaneously chooses one orientation, giving rise to three-state Potts nematicity, which is qualitatively distinct from the commonly observed Ising nematic order in multi-orbital superconductors. 
In the dynamical evolution, the lifetime of the Potts-nematic state is around $20$ ms. The emergence and disappearance of  the Potts-nematic order  in dynamics are found to coincide with  the atomic phase coherence in the excited-band. 
Our theory analysis shows that the Potts-nematic symmetry breaking is captured by an orbital-$sp^2$ (with $s$, $p_x$, and $p_y$ hybridized) lattice model (see Fig.~\ref{fig:f1}{\bf b})~\cite{2012_Sun_NatPhys,2013_Li_NatComm,2014_BoLiu_NatComm}, yet with strong many-body renormalization effects caused by single-particle inter-orbital mixing between $p_x$ and $p_y$. 
This effect is  absent in the square lattice~\cite{2013_Liu_PRA} but unavoidable in the hexagonal lattice, which makes the $p_x$-$p_y$ orbital Josephson coupling generically renormalize from the positive to the negative side in our field theory analysis. This work opens up a wide window to explore rich correlated vestigial orders in orbital-mixed atomic superfluids\cite{2006_Wu_PRL,2008_Wu_PRL,2008_Liu_PRL,2012_Sun_NatPhys,2013_Li_NatComm,2014_BoLiu_NatComm,2016_XuLiu_PRL} .

Our experiment is based on a $^{87}$Rb BEC with $10^5$ atoms in a quasi 2D hexagonal optical lattice, composed of two classes of tube-shaped lattice sites, denoted as ${\cal A}$ and ${\cal B}$ (see Fig.~\ref{fig:f1}).  
Atoms are confined in $800$ tubes, with each tube containing $60$ atoms on average. The temperature of atoms before loading into the optical lattice is 75nK, for which about $70\%$ of the atoms are condensed in our experiment.

The lattice potential is formed by three intersecting far-red-detuned laser beams in the x-y plane with an enclosing angle of 120$^{\circ}$.   Each beam is formed by combining two linearly polarized light with polarization directions oriented in the x-y plane (denoted as in-light) and along the z-axis (denoted as out-light), respectively. The in- and out-light form an inversion symmetric honeycomb lattice, and a simple triangular lattice, respectively, whose lattice strengths ($V_{\rm in}$ and $V_{\rm out}$)  are separately tunable. The out-to-in light intensity ratio is denoted as $\tan^2 \alpha = V_{\rm out} /V_{\rm in}$. The well-depths at $\mathcal{A}$ and $\mathcal{B}$ sites are made different by aligning two lattices in a way that $\mathcal{A}$ ($\mathcal{B}$) sites of the honeycomb lattice are enhanced (weakened) by the potential minima (maxima) of the triangular lattice or the other way around, which is controllable by choosing relative phases between the in- and out-light, denoted as $\theta_{1,2,3}$~\cite{supplement}.
The lattice has a similar geometry as in previous experiments~\cite{PhysRevLett.70.2249,Becker_2010,SoltanRN158}.  
We first  adiabatically load BEC into the ground band optical lattice. 
 The phase differences are initially set to be $\theta_{1,2,3}=(2\pi/3,4\pi/3,0)$, for which ${\cal B}$ sites are deeper than the ${\cal A}$ sites.  The ground state BEC forms at the $\Gamma$ point, which respects all lattice symmetries.
In real space atoms mainly reside in the $s$-orbitals of ${\cal B}$ sites. 
We then adopt the projection protocol developed for loading atoms to excited bands of a square lattice~\cite{2011_Hemmerich_NatPhys}. 
We switch the phase differences $\theta_{1,2,3}$ rapidly (within $0.1$ ms) to the reverse case with $\theta_{1,2,3}=(4\pi/3, 2\pi/3, 0)$, making $\mathcal{A}$ sites much lower than $\mathcal{B}$. 
{In this way the atomic sample is directly projected onto the excited band.}  
By selecting an appropriate combination of laser intensities having $V_{\rm in} + V_{\rm out} = 30 E_R$ ($E_R$ the single-photon recoil energy), and $\alpha = 14^\circ$,  a second-band population-ratio of    $50\%$ is achieved, as measured by band mapping techniques (Fig.~\ref{fig:f1}).
In this work, we choose laser intensity  such that  $s$-orbital of ${\cal B}$ sites are near resonance with $p_{x,y}$-orbitals of ${\cal A}$ sites in the final lattice, and consequently the second, third, and forth bands are close-by in energy~\cite{supplement}.

The quantum tunnelings at the final stage is then  described by an $sp^2$-orbital-hybridized model,
\bea
 H_0 &=&
 t_{sp} \sum_{{\bf r} \in {\cal B}}  \sum_{a=1,2,3} \left[ \hat{s}_{\bf r} ^\dag ( \vec{\hat{ p}}_{{\bf r}+{\bf d}_a} \cdot {\bf e}_a) +H.c.\right] \nn \\
&& -\mu_s \sum_{{\bf r} \in {\cal B}} \hat{s}_{\bf r}^\dag \hat{s}_{\bf r}
 -\mu_p \sum_{{\bf r}' \in {\cal A}} \vec{\hat{p}}\,^\dag_{{\bf r}'} \cdot \vec{ \hat{p}}_{{\bf r}'},
\label{eq:H0}
\eea
Here, $\hat{s}$ and $\hat{p}$ represent quantum mechanical annihilation operators for $s$- and $p$-orbitals, and the shorthand notation $\vec{\hat{p}} = (\hat{p}_x, \hat{p}_y)$.
The unit vectors ${\bf e}_1 = (-1,0)$, ${\bf e}_2 = (1/2, -{\sqrt{3}}/{2} )$, and ${\bf e}_3 = (1/2, {\sqrt{3}}/{2} )$
and corresponding ${\bf d}_a=(2\lambda/3\sqrt{3}){\bf e}_a$ mark the relative position between the two sub-lattices (Fig.~\ref{fig:f1}), 
with $\lambda$ the laser wavelength. The quantum tunneling between ${\cal A}$ and ${\cal B}$ sub-lattices is $t_{sp}$, which is about $2\pi \times 540$ Hz in our experiment. The chemical potentials for $s$- and $p$-orbitals are denoted as $\mu_s$, and $\mu_p$, respectively. 
The many-body quantum effects are modeled by the $s$-orbital interaction,
$H_{\rm int, s}  =U_s /2\sum_{ {\bf r}\in {\cal B}} \hat{s}_{\bf r} ^\dag \hat{s}_{\bf r}^\dag \hat{s}_{\bf r} \hat{s}_{\bf r}$, and
the $p$-orbital interaction,
\bea
\label{eq:Hpint} 
H_{\rm int, p}
&=&
  \sum_{ {\bf r }\in {\cal A} }
  \left\{ J  \left[ \hat{p}_{x, {\bf r}} ^\dag \hat{p}_{x,{\bf r}} ^\dag \hat{p}_{y, {\bf r}}  \hat{p}_{y, {\bf r}}+ H.c. \right] \right\}   \\
&+& \sum_{ {\bf r }\in {\cal A} }
 \left\{ \frac{1}{2} \sum_{\alpha,\beta \in \{x,y\}}U_{p,\alpha \beta} \hat{p}_{\alpha, {\bf r}} ^\dag \hat{p}_{\beta,{\bf r}} ^\dag\hat{p}_{\beta, {\bf r}}  \hat{p}_{\alpha, {\bf r}}
 	\right\}. \nn
\eea

In the language of group theory, $s$-orbital transforms according to a one-dimensional representation of the  lattice symmetry group $C_{3v}$ (A$_1$),
and $p$-orbitals correspond to
the two-dimensional representation (E).
The $p$-orbital couplings are constrained by $U_{p,xx} = U_{p,yy} \equiv U_{p\parallel }$, $U_{p,xy} = U_{p,yx} \equiv U_{p\perp}$, $J = (U_{p \parallel}-U_{p \perp} )/2$, according to symmetry analysis. 
In our experiment, the density interactions $U_s$, $U_{p\parallel}$, and $U_{p \perp}$ are found to be comparable with the tunneling $t_{sp}$, and the Josephson coupling $J$ is one order of magnitude smaller~\cite{supplement}. 
{By loading cold atoms into the excited band in our hexagonal lattice,} 
a quantum many-body system with $sp^2$-orbital hybridization is achieved, which is a versatile platform to host rich physics such as large-gap topological phases~\cite{2009_Xia_NatPhys,2014_Wu_PRB}, exotic orbital frustration~\cite{2008_Liu_PRL,2008_Wu_PRL}, and novel carbon structure~\cite{2016_Dresselhaus_carbon} analogies.

\begin{figure}[htp]
\centerline{\includegraphics[angle=0,width=.5\textwidth]{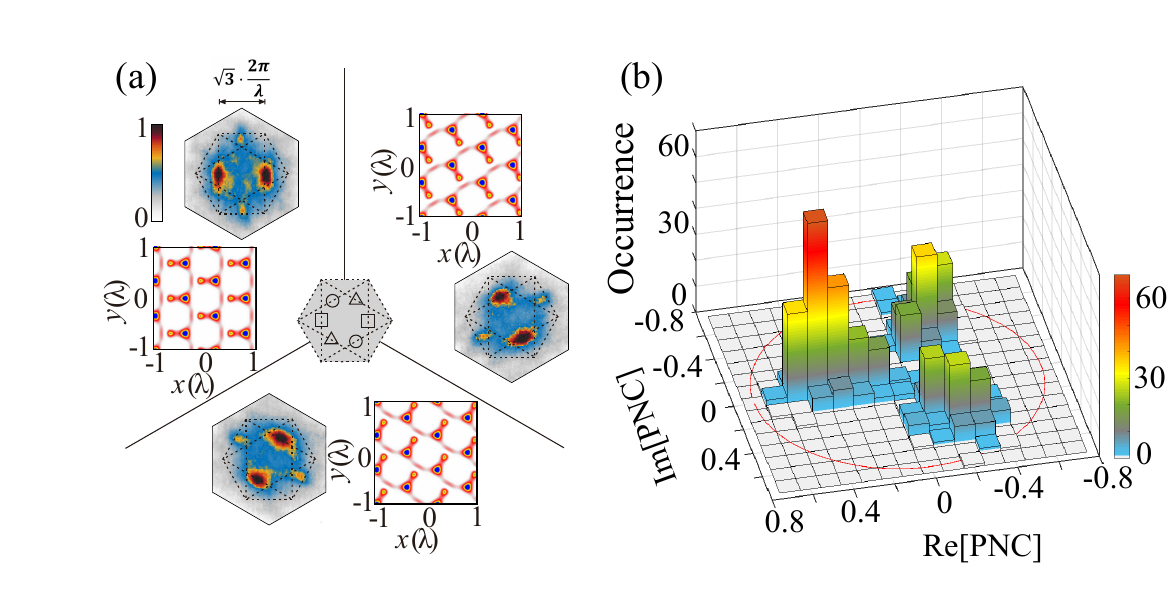}}
\caption{
{\bf Spontaneous Potts-nematic symmetry breaking in the hexagonal optical lattice}.
{\bf a}, the averaged momentum distribution. 
{Atoms loaded in the excited band are found to spontaneously accumulate at one of the three $M$ points.}  
We introduce a Potts nematic contrast (PNC in Eq.~\eqref{eq:PNC}),
where $n_{\square, \bigcirc, \bigtriangleup}$ correspond to momentum distributions in three separate regions as marked in the middle of {\bf a}, related to each other by a three-fold lattice rotational symmetry. 
In the three panels, we first classify the experimental images into three classes according to the polar angle of the nematic contrast ${\rm arg} ({\rm PNC}) \in (-\pi/3, \pi/3)$, $ (\pi/3, \pi)$, or $(\pi, 4\pi/3)$, and then take the average within each class.
In the panel insets parameterized by $x(\lambda)$ and $y(\lambda)$, 
{we show the real space density extracted from the Bloch functions at the $M$ points,} 
which shows a bond order  that breaks the lattice rotation symmetry in real space.
{\bf b}, the statistical occurrence of the nematic contrast. The nematic contrast extracted from the experimental data shows the spontaneous breaking of the three-fold lattice rotation symmetry, i.e., the emergence of the Potts nematic order in this quantum many-body system.
}
\label{fig:f2}
\end{figure}

{Right after the lattice switch we have cold atoms  reside symmetrically on the $\Gamma$ point of the second band.} 
We then hold the system for $5$ ms,
and  take the measurements of momentum distribution of the system through time-of-flight (TOF). We repeat the same experiment for $600$ times, and  then perform statistics on the independently obtained TOF images. The results are shown in Fig.~\ref{fig:f2}.
To diagnose the Potts-nematic order, we divide the momentum space into three regions marked as $\square$, $\bigcirc$, and $\bigtriangleup$, related to each other by a $C_3$ rotation (see Fig.~\ref{fig:f2}{\bf a}). The total population in these three different regions are denoted as $n_{\square}$, $n_{\bigcirc}$, and $n_\bigtriangleup$, correspondingly. We define a complex valued Potts nematic contrast (PNC) as
\be
\label{eq:PNC}
{\rm PNC} = \frac{n_{\square}+ e^{i2\pi/3} n_{\bigcirc} + e^{i4\pi/3} n_\bigtriangleup}{n_{\square}+ n_{\bigcirc} + n_\bigtriangleup},
\ee
which vanishes only when the $C_3$ symmetry is unbroken. When the symmetry is completely broken, PNC takes discrete values from $(1, e^{i2\pi/3}, e^{i4\pi/3} )$. The occurrence of PNC collected from consecutive experimental runs (Fig.~\ref{fig:f2}{\bf b}) explicitly shows 
{that the atomic  quantum state  randomly acquires one of the three orientations.} 
The occurrence probability in the three orientations is approximately equal, with the slight difference caused by experimental imperfection. 
For example, a gradient magnetic field is added along the gravitational direction to compensate the earth gravity. 
{One of three laser beams (the one along the gravitational direction) forming the hexagonal lattice is linearly polarized while the other two are elliptically polarized. The laser beams then have different degree of fluctuations. 
The slight asymmetry observed in the distribution of Potts-nematic order is attributed to the imperfect equivalence among the three directions. 
We expect that switching to a lattice  perpendicular to the gravitational direction could improve the symmetry of the distribution, which is not carried out here due to technical limitations in our experiment.} 

We then divide the experimental TOF images into three classes according to their PNC values, and then take the average within each class. The post-classification averaged results are shown in Fig.~\ref{fig:f2}{\bf a}. 
{It is evident that atoms spontaneously accumulate the $M$ points and develop phase coherence in the excited band.} 
The kinetic energy decrease in the lattice is expected to be absorbed by the continuous degrees of freedom along the tube. 
From these results, the Bragg-peaks of the momentum distribution form a reciprocal lattice of the hexagonal lattice, which means the lattice translation symmetry remains unbroken. We thus conclude that the observed quantum state has a Potts-nematic order.
{The coherent Bragg peaks suggest the system has superfluidity~\cite{2002_Bloch_Nature}.}

\begin{figure}[htp]
\centerline{\includegraphics[angle=0,width=.5\textwidth]{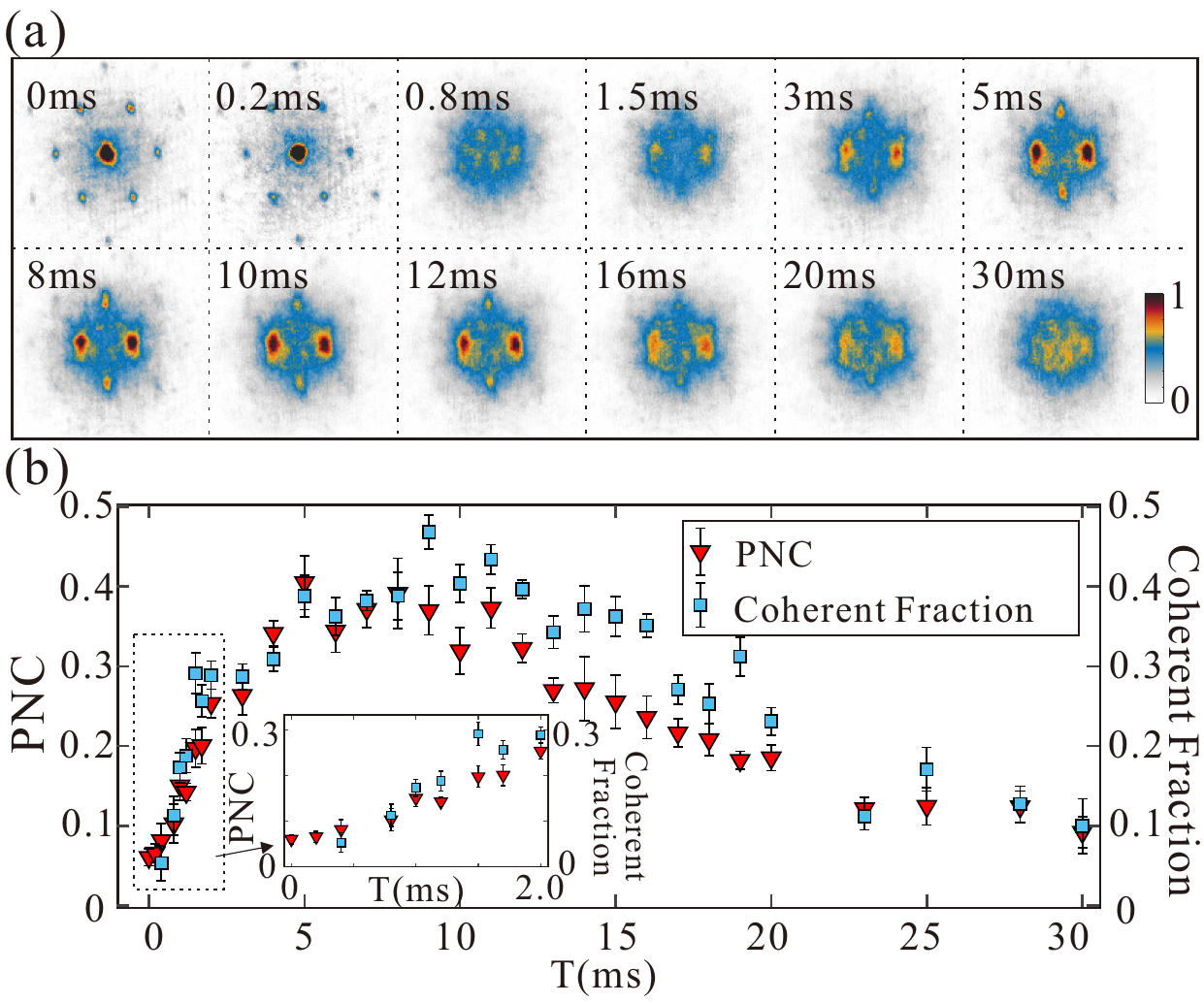}}
\caption{
{\bf Dynamical emergence of the Potts-nematic order}.
{\bf a}, Time evolution of  momentum distribution. In {\bf a},
{we average over the experimental results having a Potts nematic contrast (PNC) with ${\rm arg} ({\rm PNC}) \in (-\pi/3, \pi/3)$.}
{\bf b}, Evolution of the PNC and the coherent fraction~\cite{supplement}.
The timepoint we quench the lattice (see Fig.~\ref{fig:f1}) is set to be $0$ in this plot. 
The phase coherence in the second band does not immediately form after the quench but instead appears about several milliseconds  later. 
The emergence of Potts-nematic order coincides  with the second band phase coherence. 
The rise and disappearance of the Potts nematic order define three qualitatively distinct regimes in the quantum dynamics.
Here we average over ten experimental images at each time point, with the error bar denoting the statistical error.
}
\label{fig:f3}
\end{figure}

Since the observed  Potts-nematicity occurs in the excited band, it has finite lifetime and eventually decays in the dynamical evolution. In Fig.~\ref{fig:f3}, we show the rise and disappearance of the Potts-nematic order in the quantum dynamics. The observation implies three different stages of dynamical evolution. 
{At the first stage right after atoms are loaded to the excited band,} 
the effective mass is negative at the $\Gamma$ point causing strong dynamical instability~\cite{2008_Pethick_BEC},
which immediately (within $1$ ms) destroys the phase coherence in the lattice directions. Around $1$ ms, the momentum distribution of the atoms has no sharp features (see Fig.~\ref{fig:f3}{\bf a}). 
{At a second stage, the atomic phase coherence starts to rebuild in the excited band around several milliseconds after getting excited, 
and the Potts-nematic order emerges simultaneously. The coherent Potts-nematic quantum state remains stable up to about $20$ ms.}
The intermediate-time nematic order defines 
three distinctive regimes in quantum dynamics separated by 
the occurrence and disappearance of the spontaneous rotation symmetry breaking. Similar transient dynamics has also been found in the bipartite square lattice for a chiral Bose-Einstein condensate~\cite{2011_Hemmerich_NatPhys}. 
We expect the relatively long lifetime of the transient many-body state compared to the band relaxation time to be captured by a quantum Boltzmann equation~\cite{2020_Mueller_PRA}.

\begin{figure}[htp]
\centerline{\includegraphics[angle=0,width=.5\textwidth]{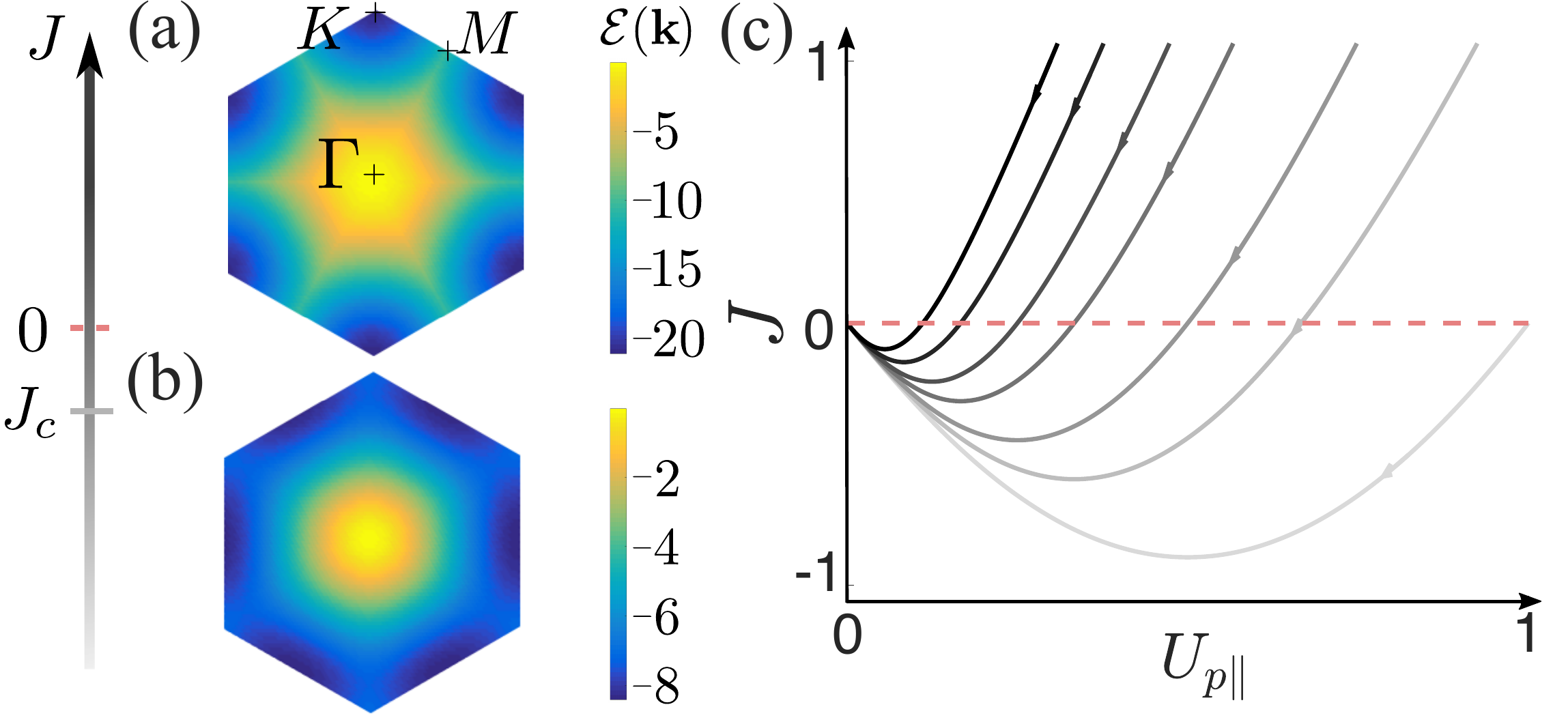}}
\caption{
{\bf Theoretical quantum phase transitions with varying the orbital Josephson coupling.}
 The orbital Josephson coupling $J$ is introduced in Eq.~\eqref{eq:Hpint}.
{\bf a}, The Gross-Pitaevskii energy ${\cal E} ({\bf k})$ for a plane-wave condensate at a lattice momentum ${\bf k}$. Here we choose $t_{sp}$ as an energy unit. 
The chemical potentials are set at $\mu_s /t_{sp}  =0.1$, $\mu_p =0$, 
the interaction strengths are $U_s/t_{sp} = U_{p,\parallel}/t_{sp} = 0.5$, $J/t_{sp} = 0.4$ and $-0.4$ in {\bf a} and {\bf b}, respectively, and $U_{p,\perp}$ is fixed respecting the lattice rotation symmetry.  The energy ${\cal E} ({\bf k})$ has minima at $K$ ($M$) points for $J >J_c$ ($J<J_c$). 
The ground state condensates are chiral and Potts-nematic correspondingly.
{\bf c}, The sketch of the renormalization of the $p$-orbital couplings to low energy. 
The multiple curves correspond to different choice of bare couplings.
The feature of $J$ renormalizing to the negative side is generic for the hexagonal lattice. 
In {\bf c}, the couplings are in arbitrary units.
}
\label{fig:f4}
\end{figure}

To gain insight into the mechanism supporting the Potts-nematic order in the $sp^2$-orbital hybridized band, 
we provide a mean field theory analysis assuming a plane-wave condensate. Taking a trial condensate wavefunction with $\langle s_{\bf r}\rangle  = \phi_s e^{i{\bf k}\cdot {\bf r}} $, $\langle p_{x,y,{\bf r}} \rangle = {\phi_ {x,y}} e^{i{\bf k}\cdot {\bf r}} $, with $\phi_s$, $\phi_{x,y}$ the variational parameters. For each lattice momentum ${\bf k}$ we minimize the energy by varying $\phi_{s,x,y}$, and the resultant energy is denoted as ${\cal E}({\bf k})$ and shown in Fig.~\ref{fig:f4}.
With the  orbital Josephson coupling $J>0$ (Eq.~\eqref{eq:Hpint}), both the kinetic and interaction energies favor a condensate at $K$ points which breaks the time-reversal symmetry but respects the rotation symmetry. The corresponding condensate has a $p_x+ip_y$ character as in the square lattice~\cite{2011_Hemmerich_NatPhys,2016_Li_RPP}. With the Josephson coupling $J<0$, minimizing the kinetic and the interaction energies meet frustration, as interaction then favors  $p$-orbital polarization. Once the Josephson coupling is beyond a certain threshold $J<J_c \sim(-t_{sp})<0$, the competition between kinetic and interaction energies leads to a condensate at $M$ points, breaking the lattice rotation symmetry. It is worth noting here that at the field theory tree level~\cite{1994_Shankar_RMP}, $J$ is always positive for repulsive atoms. The observation of the Potts-nematic order in the experiment is thus beyond the simple mean field theory and requires considering renormalization effects. 
Integrating out higher momentum modes, the interaction strengths among the low-energy modes renormalize as 
$ {\Delta U_s } \propto -[U_s (\Lambda)] ^2$, 
$\Delta [U_{p\parallel} + 2J] \propto - [U_{p\parallel} (\Lambda) + 2J (\Lambda)] ^2$, 
$\Delta U_{p \perp}  \propto -[ U _{p\perp} (\Lambda) ]^2$~\cite{supplement}. 
We find that the coupling $J$ generically renormalizes to the negative side in our system due to the single-particle orbital mixing unavoidable in the hexagonal lattice (Fig.~\ref{fig:f4})---the mediated single-particle mixing between $p_x$ and $p_y$ on nearby ${\cal A}$-sites induced by an $s$ orbital is at the order $\hbar \times 100$ Hz according to a perturbative estimate, $t_{sp}^2/ (\mu_s - \mu_p)$. The essential difference between the renormalization of $U_\perp$ and $U_\parallel$ is that, it is diagonal for $U_\perp$ whereas it is non-diagonal for $U_\parallel$.  
The renormalization effects then stabilize the Potts-nematic order. This is in sharp contrast to the chiral $p$-orbital condensate in the square lattice~\cite{2011_Hemmerich_NatPhys,2013_Liu_PRA}, where the physics is captured within a simple mean field theory in absence of $p_x$-$p_y$ orbital mixing. The many-body renormalization effect caused phase transition has also been found for atoms in a multimode cavity~\cite{2017_Demler_PRA}. 
We remark here that although to fully determine whether the observed state is a condensate requires further interference measurements, our theory captures the Potts-nematic symmetry breaking regardless of the condensation, with thermal fluctuations taken into account~\cite{2013_Li_Arun_NatComm}.

{\it Conclusion and Outlook.---} 
By loading bosonic atoms into a hexagonal $sp^2$ optical lattice, we find emergence of a Potts-nematic quantum state in dynamics. The Potts-nematic order  spontaneously breaks a three-fold rotation symmetry of the lattice. Our field theory analysis shows that the Potts-nematic order is stabilized by intricate renormalization effects  caused by inter-orbital mixing.  We expect our experiment to stimulate investigation of other scenarios for the Potts nematic order as well such as thermal fluctuations, dissipative dynamics, and lattice imperfections.

{\it Acknowledgement.---}
This work is supported by National Program on Key Basic Research Project of China (Grant No. 2016YFA0301501, Grant No. 2017YFA0304204), National Natural Science Foundation of China (Grants No. 61727819, 11934002, 91736208, and 11774067, 11920101004), Natural Science Foundation of Shanghai City (Grant No. 19ZR1471500), and
 Shanghai Municipal Science and Technology Major Project (Grant No.2019SHZDZX01).

\bibliographystyle{naturemag}
\bibliography{references}




\begin{widetext}

\newpage
\renewcommand{\theequation}{S\arabic{equation}}
\renewcommand{\thesection}{S-\arabic{section}}
\renewcommand{\thefigure}{S\arabic{figure}}
\renewcommand{\thetable}{S\arabic{table}}
\setcounter{equation}{0}
\setcounter{figure}{0}
\setcounter{table}{0}

\begin{center} 
{\Huge Supplementary Material} \\
\end{center}

\begin{center}
\section{Experimental details}
\end{center}

\subsection{Creation of the controllable hexagonal lattice}
The lattice potential is formed by three intersecting red-detuned laser beams in the x-y plane with an enclosing angle of  $120^\circ$
[\textit{see Fig.~1 in the main text}]. 
{The relative orientation of the laser beams with respect to the magneto-optical trap and the gravitational direction in our experiment  is illustrated in Fig.~\ref{fig:SuppExpSetup}.} 
Each laser beam is formed by combining two linearly polarized light with polarization directions oriented in the lattice plane (denoted as in-light) and along the z-axis (denoted as out-light), respectively. The electric field experienced by the atoms is given by
\begin{equation}
	{\bf E}({\bf r}, t)=
		E_{\text{out}} \sum_{j=1,2,3} {\bf e}_z \cos{\left({\bf k}_j \cdot {\bf r}-\theta_{j,\text{out}} - \omega t\right)}
		+E_{\text{in}} \sum_{j=1,2,3} (\hat{{\bf k}}_j \times {\bf e}_z) \cos{\left({\bf k}_j \cdot {\bf r}-\theta_{j,\text{in}} - \omega t\right)},
\end{equation}
where we have ${\bf k}_1 = (\sqrt{3}\pi, -\pi)/\lambda$, ${\bf k}_2 = (-\sqrt{3}\pi, -\pi)/\lambda$, ${\bf k}_3 = (0, 2\pi)/\lambda$,  $\hat{{\bf k}}_j = {\bf k}_j/\left\vert{\bf k}_j\right\vert$, and $E_{j,\text{in} (\text{out})}$ and $\theta_{j,\text{in} (\text{out})}$ are the electric field amplitude and  the phase of the in-light (out-light) of each beam. The corresponding laser intensity is then $I({\bf r}) = \overline{ \left\vert {\bf E}({\bf r}) \right\vert^2 }$, given as
\bea
I({\bf r})
	& =& I_0 +
		I_{\text{out}}
			\sum_{\langle i,j \rangle}\cos\left[({\bf k}_i-{\bf k}_j) \cdot {\bf r} - (\theta_{i,\text{out}}-\theta_{j,\text{out}})\right]  \nn \\
		&-&\frac{1}{2}I_{\text{in}}
			\sum_{\langle i,j \rangle}\cos\left[({\bf k}_i-{\bf k}_j)\cdot {\bf r} - (\theta_{i,\text{in}}-\theta_{j,\text{in}})\right],
\eea
where $\overline{ \left\vert {\bf E}({\bf r}) \right\vert^2 }$ denotes the time average, $I_0=3\left( E_{\text{out}}^2 + E_{\text{in}}^2 \right)/2$, $I_{\text{out}(\text{in})}=E_{\text{out}(\text{in})}^2$ and the summation is restricted to $\langle 1,2 \rangle$, $\langle 2,3 \rangle$ and $\langle 3,1 \rangle$. With large red-detuning in our experiment, the resultant optical potential on atoms takes the form
\bea
	V({\bf r}) &=&
		-V_{\text{out}}
			\sum_{\langle i,j \rangle}\cos\left[({\bf k}_i-{\bf k}_j) \cdot {\bf r} - (\theta_{i,\text{out}}-\theta_{j,\text{out}})\right] \nn \\
		&+&\frac{1}{2}V_{\text{in}}
			\sum_{\langle i,j \rangle}\cos\left[({\bf k}_i-{\bf k}_j)\cdot {\bf r} - (\theta_{i,\text{in}}-\theta_{j,\text{in}})\right].
\eea
By choosing a convenient set of coordinates, the optical potential further simplifies to
\begin{eqnarray}
	V({\bf r}) &=&-V_{\text{out}}\sum_{\langle i,j \rangle}
	\cos\left[({\bf k}_i-{\bf k}_j) \cdot {\bf r} +(\theta_{j, {\rm out}} - \theta_{j, {\rm in}})  - (\theta_{i, {\rm out}} -\theta_{i, {\rm in}} )\right] \nn \\
	&+&\frac{1}{2}V_{\text{in}}\sum_{\langle i,j \rangle}\cos\left[({\bf k}_i-{\bf k}_j)\cdot {\bf r}\right].
\label{eq:SVr}
\end{eqnarray}
The $V_{\text{out}}$ and $V_{\text{in}}$ terms correspond to a simple triangular lattice and an inversion symmetric honeycomb lattice, respectively.
It is worth remarking here that the alignment of these two only depend on the relative phases between the two polarization directions within each laser beam, which is stabilized using a feedback control in our experiment (Section~\ref{sec:stabilization}).
With this optical potential, the relative position of the two lattices is controllable by tuning the phases $\theta_j \equiv \theta_{j, {\rm out}} -\theta_{j, {\rm in}}$.
Given the cyclic constraint $(\theta_1 -\theta_2) + (\theta_2 - \theta_3)  + (\theta_3 - \theta_1)=0$,  we have two independent degrees of freedom from which the two-dimensional relative position between the triangular and the honeycomb lattices is tunable to arbitrary degree.
 In the experiment we set $\theta_3 =0$ for simplicity, yet without compromise of the lattice controllability.  In order to make $\mathcal{A}$ sites deeper than $\mathcal{B}$,
 we choose $\left\{ \theta_1,\theta_2,\theta_3 \right\} = \left\{ 4\pi/3,2\pi/3,0 \right\}$, for which the potential minima (maxima) of the triangular lattice locates at the $\mathcal{A}$ (${\cal B}$) sites of the honeycomb lattice.
 This is reversed with $\left\{ \theta_1,\theta_2,\theta_3 \right\} = \left\{ 2\pi/3,4\pi/3,0 \right\}$.
A fast swap between two configurations can be achieved within 0.1ms in our experiment.

The laser intensities of the in- and out-light are separately controllable, and the resultant potential strengths are denoted by $V_{\rm in}$ and $V_{\rm out}$, whose ratio
$\tan^2\alpha=V_{\text{out}}/V_{\text{in}}$ has been introduced in the main text to describe the relative intensity.
 In the experiment, we choose $V_{\rm in} + V_{\rm out}$ to be thirty times of photon-recoil-energy, $E_r$, for which $s$-orbitals on ${\cal B}$-sites are near resonance with $p$-orbitals on ${\cal A} $ sites in the final lattice configuration.
The laser intensities are carefully stabilized to avoid lattice potential deformation in the experiment to be explained below.

\begin{figure}[htp]
\centerline{\includegraphics[angle=0,width=.5\textwidth]{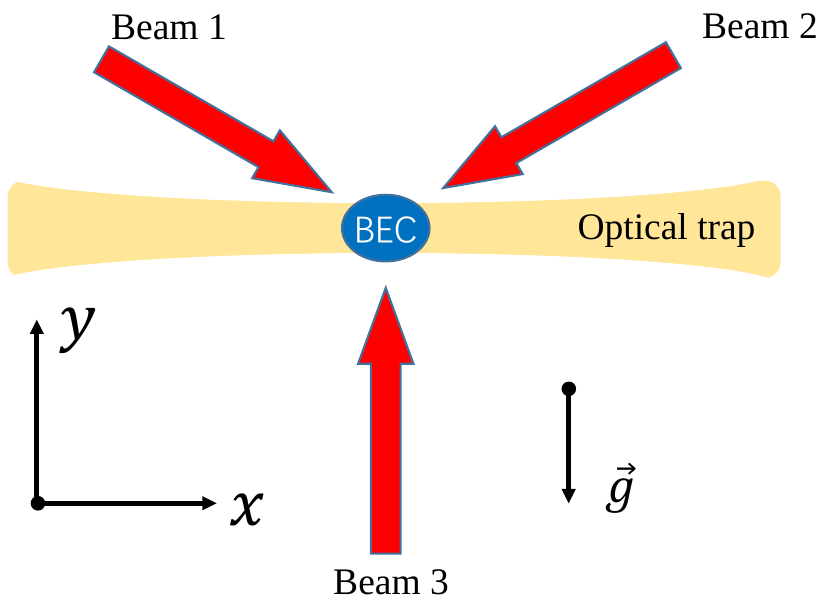}}
\caption{Illustration of directionality of the magneto-optical trap and hexagonal optical lattice. The magneto-optical trap includes a 1064nm laser beam along the x-direction, providing an optical dipole trap, and a gradient magnetic field to compensate earth-gravity ($\vec{g}$). The three laser beams (Beam $1$, $2$, $3$) forming the optical lattice are shown with red arrows. Beam $3$ is linearly polarized and is  along the gravitational direction. Beam $1$ and $2$ are elliptically polarized.  
 }
\label{fig:SuppExpSetup}
\end{figure}

\begin{figure}[htp]
\centerline{\includegraphics[angle=0,width=\textwidth]{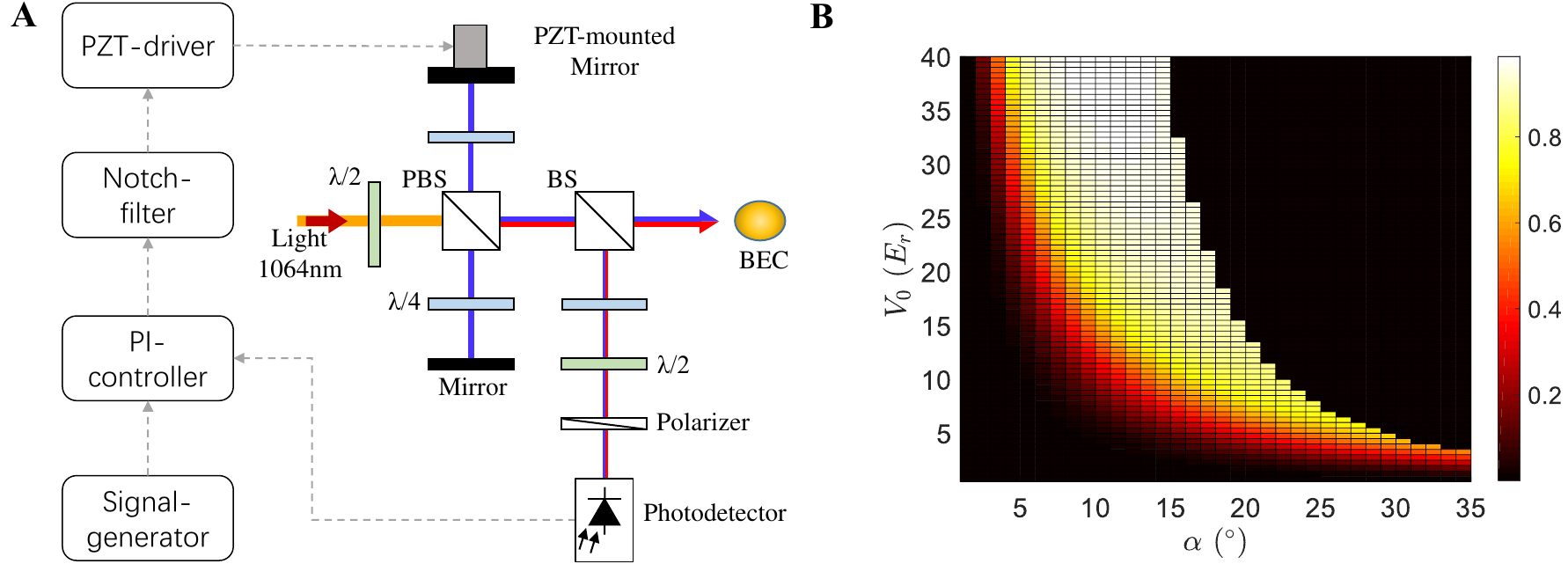}}
\caption{
{\bf A},
Schematic of the phase control system. An inclined linear polarized beam is incident into a polarization-beam-splitter (PBS) where it splits into an in-light beam (red, building an inversion symmetric honeycomb lattice) and an out-light beam (blue, building a simple triangular lattice). The out-light beam goes through an additional optical path, which is controlled by a piezoelectric (PZT)  mounted mirror. After the in- and out-light beams recombine we have an elliptically polarized light.
A small fraction of the elliptically polarized light is split out by using a beam splitter (BS) for phase-stabilization purpose. With a half-wave plate and a polarizer, the relative phase $\theta_{1,2,3}$ is reflected by the intensity collected by the photodetector.
We utilize a proportion-integral (PI) controller and a Notch-filter to build a feedback control that stabilizes $\theta_{1,2,3}$  to the desired value.
{\bf B},
The wavefunction overlap $\eta$ (Eq.~\eqref{eq:eta}) with different lattice depth and light intensity ratio (parameterized by the angle $\alpha$).
At small $\alpha$, the overlap $\eta$ increases for larger $\alpha$. When $\alpha$ reaches a certain critical value $\alpha_c$, the second and third bands touch at the $\Gamma$ point, and $\eta$ then suddenly drops down to zero due to level crossing.
}
\label{fig:Sup}
\end{figure}

\begin{figure}[htp]
\centerline{\includegraphics[angle=0,width=.9\textwidth]{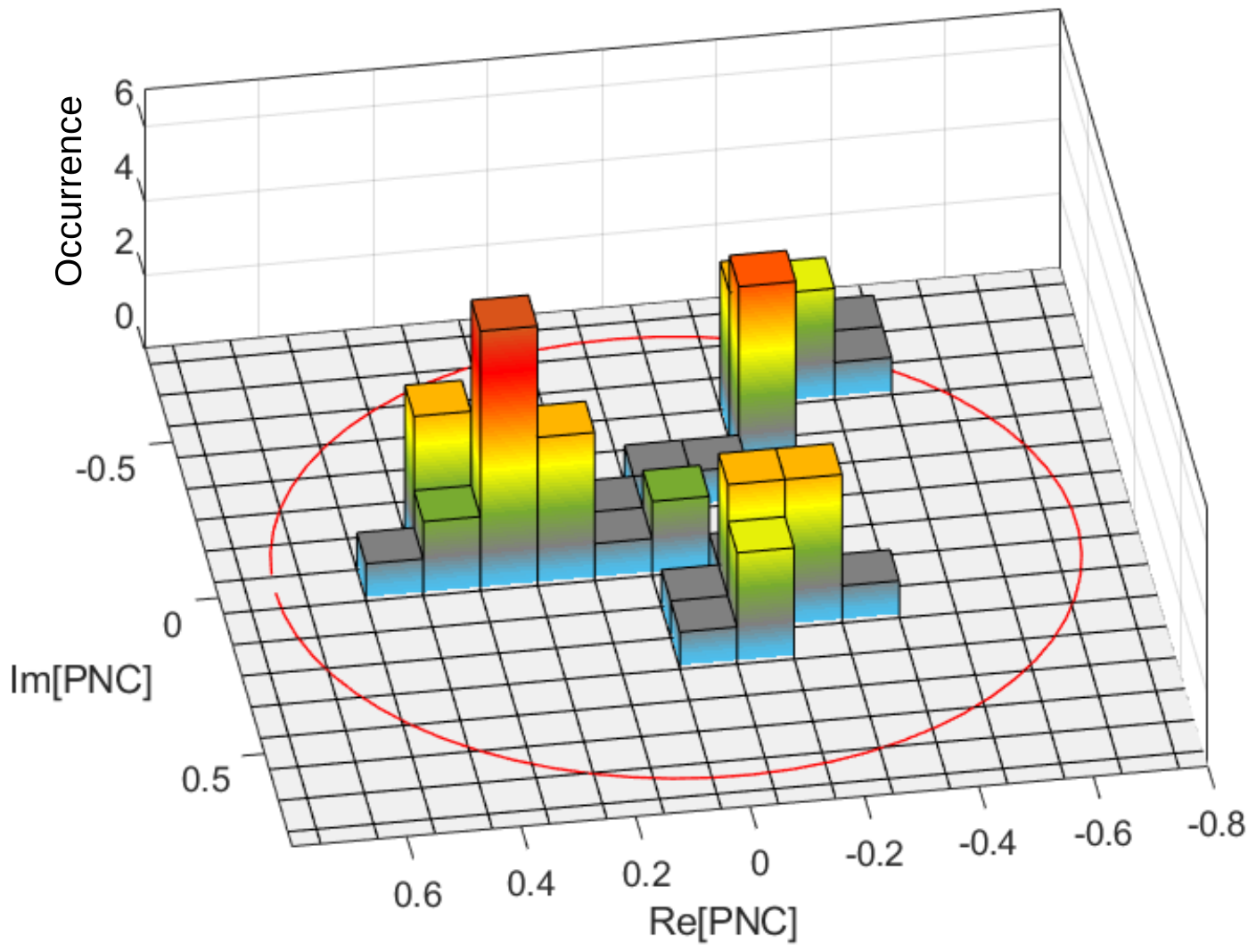}}
\caption{The statistical occurrence of the Potts nematic contrast with a lattice depth of $V_{\rm in } + V_{\rm out}$ = 25$E_r$ , whose corresponding second band population ratio is about 43\% in our experiment.
Here, the statistics is performed over $50$ consecutive experimental runs.
 }
\label{fig:SPottsNematic}
\end{figure}

\subsection{Loading and detection procedure.}
A BEC of $^{87}$Rb is prepared in a hybrid trap with the harmonic trapping frequencies $(\omega _{x},\omega _{y},\omega _{z})=2\pi \times $(28Hz, 55Hz, 60Hz). Then the optical lattice is adiabatically ramped up within 80ms. In this stage, $\mathcal{B}$ sites are chosen as the deeper ones. After holding for 1ms,
we quickly swap the depths of $\mathcal{A}$ and $\mathcal{B}$, such that $\mathcal{A}$ sites suddenly becomes deeper than ${\cal B}$.
In order to maximize the efficiency of this high-band loading protocol,
we carefully choose a combination of lattice depth $V_0=(V_{\rm in} + V_{\rm out})$ and out-to-in intensity ratio $\tan^2\alpha$ such that $s$-orbital of $\mathcal{B}$ sites are near resonance with $p_{x,y}$-orbitals of $\mathcal{A}$ sites in the final lattice. Quantitatively, we optimize the following wavefunction overlap,
\be
\eta (V_0, \alpha) \propto \int_{\textbf{unit cell}} dx dy \psi_{1,\Gamma}(x, y; V_0, \alpha) \cdot \psi_{2,\Gamma}(x, y; V_0, \alpha),
\label{eq:eta}
\ee
with $\psi_{1, \Gamma}$ ($\psi_{2,\Gamma}$) the $\Gamma$ point Bloch function associated with  the ground (first excited) band.
The dependence of $\eta$ on $V_0$ and $\alpha$ is shown in Fig.~\ref{fig:Sup}{\bf B} which provides important guidance to optimize the excited band loading efficiency in our experiment.
 By selecting appropriate lattice depths, we are able to load  50\% of the atoms to the second band, as shown in
 \textit{Fig.~1(E) in the main text} . The atomic population in different bands is measured using standard band mapping techniques.
 The second band population ratio varies as we use different lattice depths, but we confirm the robustness of  Potts-nematic order against the band population ratio (see Fig.~\ref{fig:SPottsNematic}).

After the swap process, we hold the atomic system up to tens of milliseconds  and measure the momentum distribution through time-of-flight (TOF without band mapping). The momentum distribution of the atoms is shown in
\textit{Fig.~2(A) in the main text}.
In the first $1.5$ ms, the phase coherence of atoms gradually disappears,
{and re-emerges within a few milliseconds.}
More technical details of the experimental platform have been provided in our earlier work~\cite{2019_Zhou_NPJ}.

\subsection{Feedback stabilization of relative phases}
\label{sec:stabilization}
In the experiment, we use the system shown in {Fig.~\ref{fig:Sup}{\bf A} for phase stabilization.
For each laser beam, we first split an inclined linearly polarized beam into two components, whose polarization directions are respectively along the x-y plane (in-light) and the z-direction (out-light).
The out-light goes along an extra optical path which is controlled by a piezoelectric (PZT) mounted mirror and stabilized with a proportion-integral (PI) controller system, and then combines with the in-light. In this way, an elliptically polarized laser beam with a controllable relative phase is obtained.
To reinforce the phase stability, we split out a small fraction of the elliptically polarized light before it enters the vacuum chamber, from which the relative phase is measured.  The phase error  is collected in real time for a feedback control on the PZT that controls the extra optical path added to the out-light. This forms a feedback loop for relative phase stabilization. With this feedback control, the relative phases $\theta_{1,2,3}$ are highly controllable and fast switch of the phases is achieved in a reliable way. To verify this feedback control, we also directly measure the polarization of the dominant fraction  of light before it enters the chamber, which confirms that the phase fluctuation is suppressed down to a level below $\pi/200$. With this experimental setup the phase-switching can be reached within 100 $\mu$s.

{
To confirm the residual phase fluctuation is tolerable, we also carry out the experiment setting $\theta_1$ away from $2\pi/3$ by amount of $\pi/30$, where we find the three nematic states still emerge. This means the phase stabilization achieved in the experiment is sufficient for the study of Potts nematic order.
This is further confirmed in our band dispersion calculation taking potential imperfect phase control into account (see Fig.~\ref{fig:SR1}).
}

\begin{figure}[htp]
\centerline{\includegraphics[angle=0,width=\textwidth]{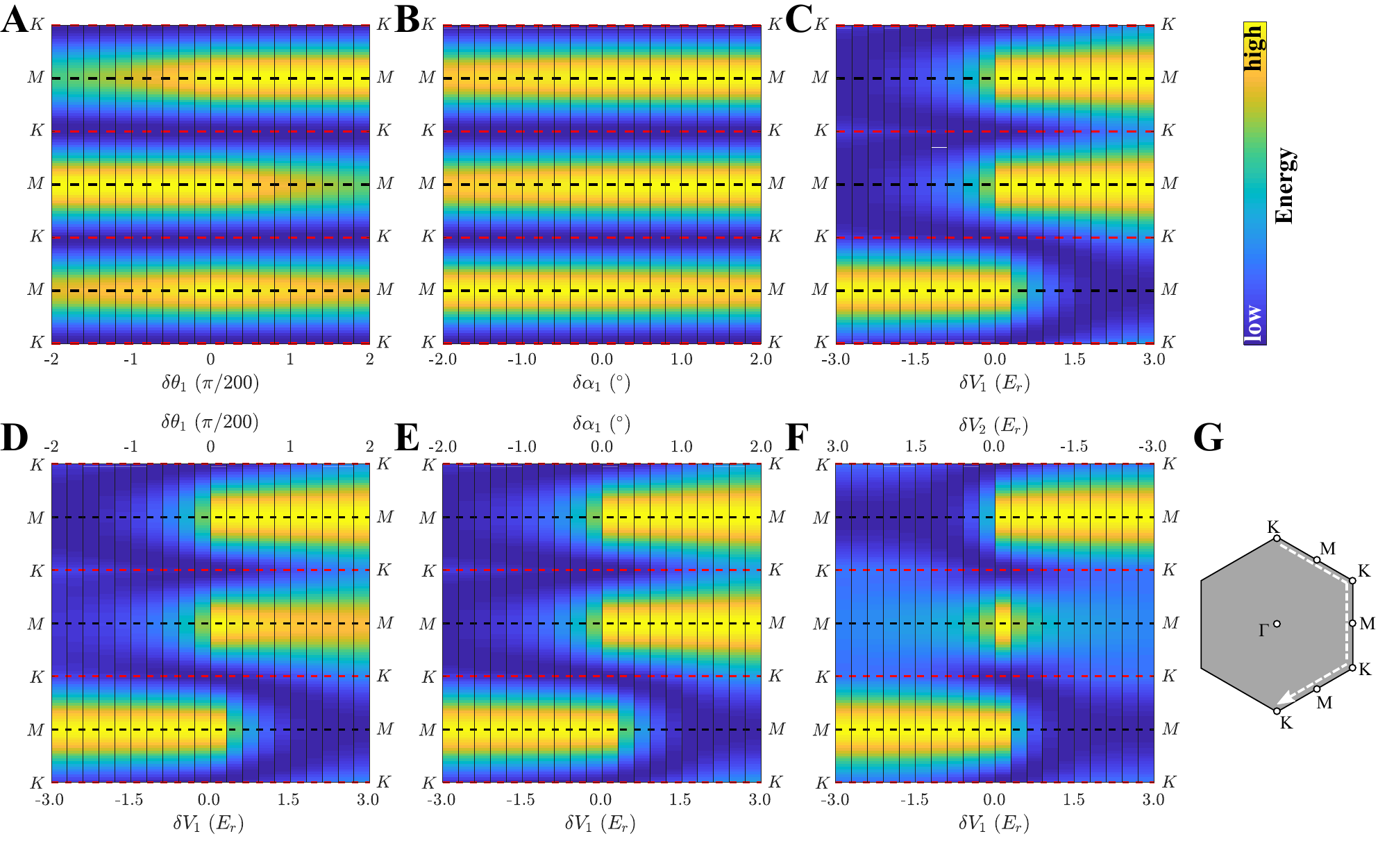}}
\caption{
Numerical results for the non-interacting dispersion of the second band of our lattice in the presence of lattice deformation. We calculate the dispersion along the line cut as shown in ({\bf G}). To illustrate the relative comparison between the energies at $K$ and $M$  points, the band dispersion is normalized. The dispersion at $K$ and $M$  points are marked by red and black dashed lines.
({\bf A}) shows the dispersion with varying the angle $\theta_1$ (Eq.~\eqref{eq:SVr}) by $\pm \pi/ 100$ from its desired value of $2\pi/3$. Although this would cause misalignment of the triangular and hexagonal lattices, the band minima still remain at $K$ points. With our feedback control method in the experiment, the imperfection of $\theta_1$ is kept well below $\pi/100$.
({\bf B}) shows the dispersion with varying the ratio $\tan ^2 \alpha_1$ between the intensities of two polarizations for one particular laser beam.  The ratio is hold fixed for the other two laser beams. As $\alpha_1$ is varied from $12^\circ$ to $16^\circ$, the band minima remain at $K$ points. And in the experiment, the imperfection in $\alpha$ depends on the extinction ratio of the half-wave plate and PBS. This extinction ratio is about $1/400$, resulting in an imperfection in $\alpha$ of about $0.3^\circ$, much smaller than $2^\circ$.
({\bf C}) shows the dispersion with varying the intensity $V_1=V_{\rm in}+V_{\rm out}$ for one of the three laser beams (Eq.~\eqref{eq:SVr}).
With the laser intensity imperfection kept below 1\% using the Kapitza-Dirac based calibration method~\cite{ovchinnikov1999diffraction,gadway2009analysis,zhou2018high} in our experiment, the band minima remain at $K$ points.  ({\bf D}, {\bf E}, {\bf F}) correspond to other potential imperfection, in which we vary two parameters together $(\theta_1, V_1)$, $(\alpha_1, V_1)$, and $(V_2, V_1)$, respectively. We confirm that the band minima cannot move from $K$ to $M$  points, given our experimental condition. Consequently, the observation of a localization of the atoms at the $M$  points of the Brillouin zone in the 2nd band could not be explained by slight deformation of the lattice geometry.
}
\label{fig:SR1}
\end{figure}

\subsection{Adjustment and calibration of lattice depth}
In order to achieve the three-fold rotation symmetry of the lattice (Fig.~1), we need to enforce the balance of laser-intensities in the three laser beams.
In the experiment, we block one of the three laser beams and adjust the other two, which then form a one-dimensional optical lattice.
{Its lattice depth is precisely determined by measuring the Kapitza-Dirac effect of the confined cold atoms~\cite{ovchinnikov1999diffraction,gadway2009analysis,zhou2018high}}.
In this way, we are able to calibrate optical imperfection in the experiment, and maintain a balance in the laser intensity among the three directions.

To confirm our controllability of the laser-intensity symmetry is sufficient in the experiment, we also deliberately make the laser intensities a bit asymmetric with a relative difference up to 5\%. This relative difference refers to the relative intensity imbalance of the three lasers forming the hexagonal lattice. 
In such experiments, the three nematic states are still observed. This implies that the residual imperfection potentially existent in the experiment  that may affect the laser intensity symmetry is negligible for the study of Potts nematic phase. This is further confirmed in our band dispersion calculation taking potential imperfect symmetry  into account (see Fig.~\ref{fig:SR1}).

{The slight systematic asymmetry observed in the distribution of the Potts-nematic order ({\it Fig.~2 in the main text}) is attributed to the imperfect equivalence among the three directions, one of which is along the earth-gravity (Fig.~\ref{fig:SuppExpSetup}). The potentially existent systematic imperfection is within the uncertainty of the experimentally adjusted parameters, which we control to our best capability.} 
We expect the distribution symmetry can be further improved by preparing a lattice that is perpendicular to the gravity direction. This has not been done in this experiment due to technical limitations in our apparatus setup.}

\begin{figure}[htp]
\centerline{\includegraphics[angle=0,width=.9\textwidth]{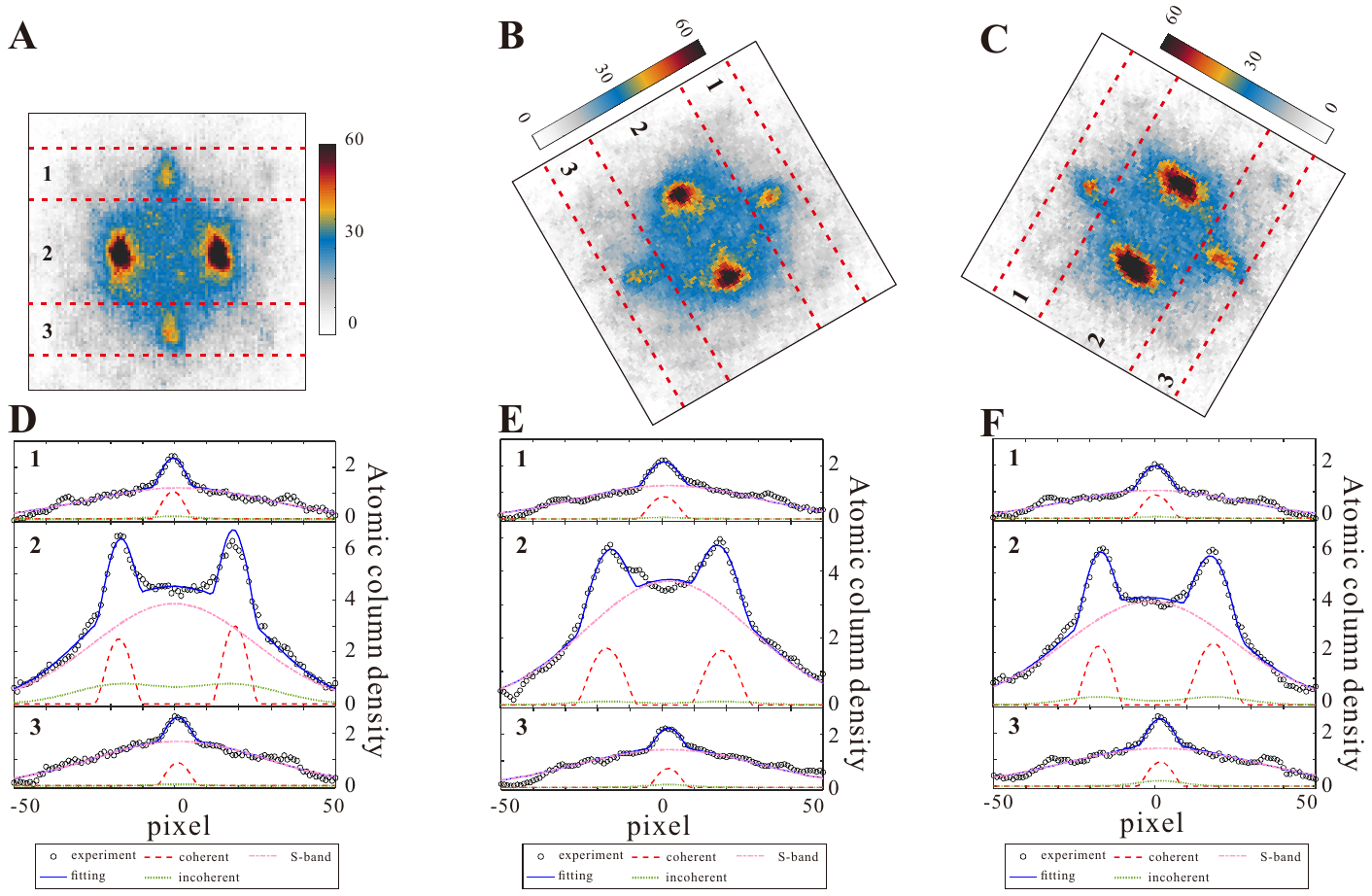}}
\caption{Extraction of the coherent fraction.
{\bf A}, {\bf B}, {\bf C}, Experimental time-of-flight images with different Potts nematic contrast ${\rm arg} ({\rm PNC}) \in (-\pi/3, \pi/3)$, $ (\pi/3, \pi)$, or $(\pi, 4\pi/3)$. We analyze the three regions of time-of-flight images as marked by $1$, $2$, and $3$ separately. The three regions are separated by red dashed lines.
{\bf D}, {\bf E}, {\bf F}, Fitting of the Bragg peaks. The black circles represent the atom column density which correspond to integration along the direction perpendicular to the dashed lines in ({\bf A}, {\bf B}, {\bf C}). The column density is then fitted to a summation of a bimodal function and a Gauss-function (shown by a pink dash dotted line), which correspond to the atom distribution in the first excited and ground bands, respectively~\cite{2011_Hemmerich_NatPhys}. The blue solid line shows the fitting. The bimodal function further involves summation of quadratic (red dashed line) and Gauss-functions (green dotted line), 
{which correspond to the coherent  and incoherent components in the excited band following the standard approach of analyzing bosonic many-body states in optical lattices.}
}
\label{fig:condensate}
\end{figure}

\subsection{Extraction of the coherent fraction}
As shown in Fig.~\ref{fig:condensate}, the time-of-flight images are divided into three regions separated by red dashed lines. In each region, we integrate the atom density distribution along the direction perpendicular to the red line, and obtain the atomic column density (indicated by black circles in the figure). 
{We adopt the approach used to analyze the excited-band phase coherence or the square lattice~\cite{2011_Hemmerich_NatPhys} and extract the coherent fraction by fitting the atom column density  to a summation of a bimodal function and a Gauss-function, which correspond to the atom distribution in the first excited and ground bands, respectively.  The number of particles contributing to the coherent component is determined from the bimodal function~\cite{2004_Esslinger_bimodal_PRL,PhysRevLett.96.180402,Becker_2010,PhysRevLett.119.100402}. The coherent fraction shown in Fig.~3 is defined to be the ratio of the number of phase-coherent atoms with respect to
the total particle number in the excited band measured through band mapping techniques
[\textit{Fig.~1 in the main text}].}

\begin{center}
\section{Theoretical analysis}
\end{center}

\subsection{Applicability of the $sp^2$ tight binding model}

In order to verify the $sp^2$ tight binding model describes the physics in our hexagonal lattice, we compare the energy dispersion of the second band derived from the $sp^2$ model with the band-structure calculation in the non-interacting regime. The tight-binding model dispersion defined by diagonaizing Eq.~(1) is shown in Fig.~\ref{fig:TBstructure}{\bf A}.
We also carry out an exact band-structure calculation by performing an expansion of the Bloch function in the plane-wave basis. The exact band-structure is shown in  Fig.~\ref{fig:TBstructure}{\bf B}.
Fig.~\ref{fig:TBstructure}{\bf C} shows the dispersion along the symmetry lines (shown in the inset) for a direct comparison of exact results with tight-binding $sp^2$ model. That the difference between the red solid line (showing the $sp^2$ model) and the blue dotted line (showing the band calculation) is below one percent, which implies that the tight-binding $sp^2$ model captures the essential physics of our lattice.

\begin{figure}[htp]
\centerline{\includegraphics[angle=0,width=\textwidth]{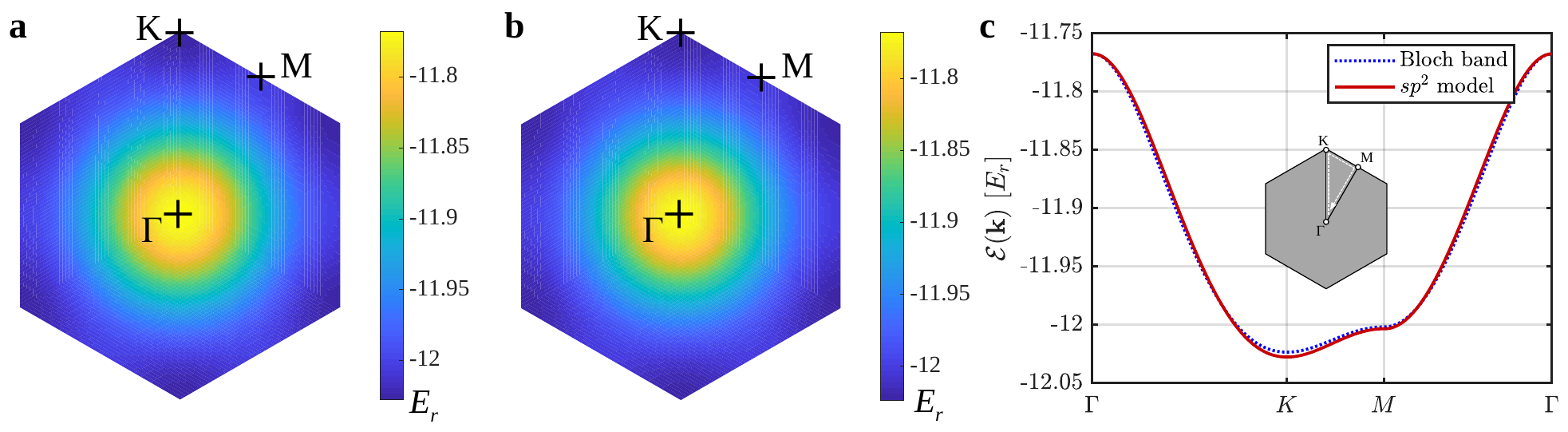}}
\caption{
{\bf A}, The band dispersion from the $sp^2$ model in the reduced Brillouin zone. {\bf B}, The result from band-structure calculation. The difference between two band structure results is below one percent, indicating that the $sp^2$ model is the model to describe the physics of our hexagonal lattice.
{\bf C} shows the dispersion along the symmetry lines (shown in the inset) for a direct comparison of exact results with tight-binding $sp^2$ model. The red solid and the blue dotted lines correspond to tight-binding and exact results, respectively.
}
\label{fig:TBstructure}
\end{figure}

\begin{figure}[htp]
\centerline{\includegraphics[angle=0,width=\textwidth]{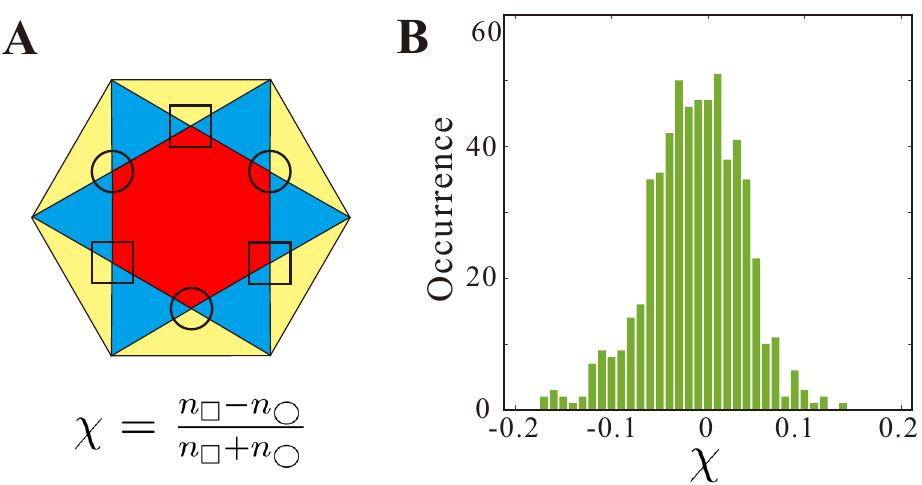}}
\caption{
{\bf Absence of chiral symmetry breaking in the experiment.}  {\bf  A}, Illustration of the Brillouin zones. Here we introduce a chiral contrast $\chi = \frac{n_\square - n_\bigcirc}{n_\square + n_\bigcirc}$, where $n_\square$ and $n_\bigcirc$ include the number of atoms near the two rotation symmetric $K$ points as shown in {\bf A}.  {\bf B}, Statistical occurrence of the chiral contrast in the $600$ experimental images whose PNC has been  shown in Fig.~2. It is evident that the BEC has the Potts nematic order rather than a chiral order.
}
\label{fig:chiralcontrast}
\end{figure}

\subsection{Mean field theory}
Although the mean field theory is imprecise, it still helps us gain insight about the underlying mechanism at phenomenological level.
The energy dispersion of the second band is derived from the $sp^2$ model
(\textit{Eq.~1 in the main text}) to be
\be
\epsilon({\bf k}) =- (\mu_s+\mu_p)/2 - \sqrt{(\mu_s-\mu_p)^2 /4+ t_{sp}^2 \left[ 3-\cos ({\bf k}\cdot {\bf a}_1) - \cos( {\bf k}\cdot {\bf a}_2) -\cos ({\bf k}\cdot {\bf a}_3) \right]  },
\label{eq:TBband}
\ee
where ${\bf a}_{1,2,3} = {\bf d}_{1,2,3} - {\bf d}_{2,3,1}$,
which has band minima at $K$ points of the hexagonal lattice.
To incorporate the interaction energy,  we take a trial condensate wavefunction with $\langle \hat{s}_{\bf r}\rangle  = \phi_s e^{i{\bf k}\cdot {\bf r}} $, $\langle \hat{p}_{x,y,{\bf r}} \rangle = {\phi_ {x,y}} e^{i{\bf k}\cdot {\bf r}} $.
Minimizing the kinetic energy would lead to a condensate at the $K$ points, and the resultant phase difference between $p_x$ and $p_y$ components is $\pm \pi/2$.
The interaction energy (per unit cell) is given by
\be
E_{\rm int} = U _s |\phi_s|^4/2 + U_{p\parallel} \left( |\phi _x|^4 + |\phi_y|^4 \right)/2 + (U_{p\parallel}-2J) |\phi _x|^2 |\phi _y|^2 + 2 J {\rm Re} [(\phi^{*} _x\phi _y)^2  ]
\ee
When the orbital Josephson coupling $J$ is positive, minimizing $E_{\rm int }$ leads to $\phi_x =  e^{\pm i\pi/2} \phi_y $. In this case, minimizing the total energy always produces a chiral condensate at $K$ points (Supplementary Information).
When the coupling $J$ is negative, minimizing the interaction leads to a phase difference of $0$, or $\pi$ between the two $p$-orbital components, which is inconsistent with chiral condensation at $K$ points.  The interaction then makes time-reversal symmetric condensates energetically more favorable. When the interaction energy dominates over the kinetic energy, the time-reversal invariant condensates at $M$ points of the hexagonal lattice  becomes the stable ground state within the $sp^2$ model. This requires $J < J_c<0$, and the critical value $J_c$ is at the order of tunneling $t_{sp}$ to compensate the kinetic energy cost.

\subsection{Ruling out  the simple mean field theory description}

In this supplementary section, we rigorously rule out the possibility of describing the experimental observation using simple mean field theory. From the experimental observation, it is evident that atoms accumulate at a single lattice momentum in the excited band.
{
In theory, condensing at a superposition of multiple lattice momenta would induce a density modulation among the lattice sites due to interference effects. This density wave order is energetically costly because the repulsive energy in our $sp^2$ orbital model favors a uniform density. We thus only consider the possibility of single lattice momentum condensation in the following.
}

In the simple mean field theory treatment, the condensate energy is obtained by  replacing the annihilation/creation operators in the Hamiltonian by their expectation values, $\langle \hat{s}_{\bf r}\rangle  = \phi_s e^{i{\bf k}_0\cdot {\bf r}} $, $\langle \hat{p}_{x,y,{\bf r}} \rangle = {\phi_ {x,y}} e^{i{\bf k}_0\cdot {\bf r}} $, where $\phi_s$, $\phi_{x,y}$, and ${\bf k}_0$ can be taken as variational parameters to minimize the mean field energy. We can re-parametrize the minimization problem by taking
\be
\begin{bmatrix}
\phi_s &=& \sqrt{n + \Delta n_{sp}/2 } e^{i (\theta + \theta_{sp}/2)}   \\
\phi_x &= &\sqrt{n - \Delta n_{sp}/4 + \Delta n_{xy} /2 } e^{i(\theta - \theta_{sp}/2 + \theta_{xy}/2)} \\
\phi_y &= &\sqrt{n - \Delta n_{sp}/4 - \Delta n_{xy} /2 } e^{i(\theta - \theta_{sp}/2 - \theta_{xy}/2)}
\end{bmatrix} .
\ee
Here $n$ corresponds to total atom number in one unit cell, $\Delta n_{sp}$ is the atom number difference between $s$- and $p$-orbitals, $\Delta n_{xy}$ is the difference between the two $p$-orbitals, $\theta_{sp}$, and $\theta_{xy}$ parameterize the relative phase among the three orbitals.
The mean field energy ($E_{\rm MF}$) contains interaction ($E_{\rm int}$) and kinetic ($E_K$) terms, i.e., $E_{\rm MF} = E_{\rm int} + E_K$. We consider a minimization procedure with $n$ and $\Delta n_{sp}$ first being fixed. From Eq.~(5,6), we find that for any choice of $n$ and $\Delta n_{sp}$, both of the kinetic energy $E_K$ and the interaction energy $E_{\rm int}$ are minimized by taking ${\bf k}_0$ to be a rotation symmetric $K$ point, $\Delta n_{xy} =0$, and $\theta_{xy} = \pi/2$, if the orbital Josephson coupling $J >0$. This means the ground state has to be a $K$ point condensate, which contradicts with experimental observation (see Fig.~2 and Fig.~\ref{fig:chiralcontrast}). And in the simple mean field theory treatment, $J$ is always positive for $^{87}$Rb atoms with repulsive interaction. 
{But the experimental observation shows atoms develop Bragg peaks at  $M$ points.} 
Therefore we rule out the possibility of using a simple mean field theory to describe our experiment.
{Based on the $M$-point condensation theory, further incorporating finite temperature thermal fluctuations is expected to first destroy the phase coherence with the discrete symmetry breaking orders surviving~\cite{2013_Li_Arun_NatComm}. }

\begin{figure}[htp]
\centerline{\includegraphics[angle=0,width=\textwidth]{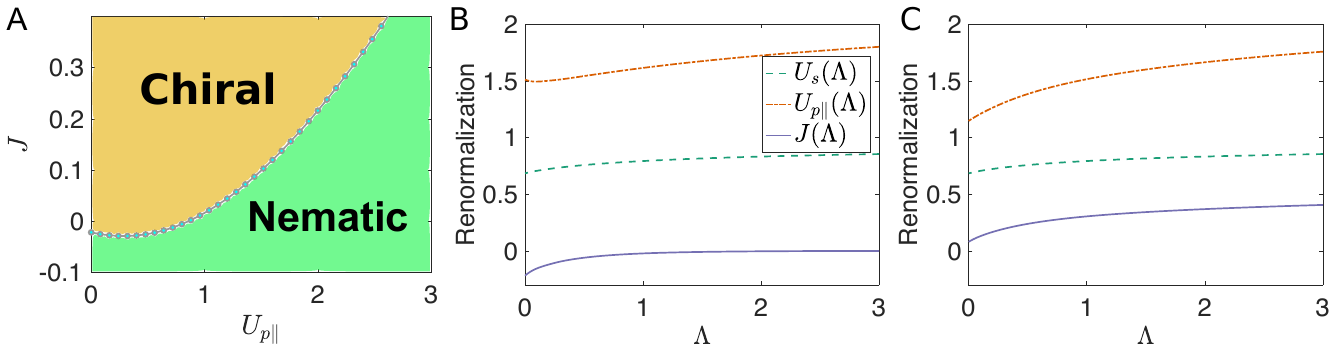}}
\caption{Phase diagram from the renormalization theory.
({\bf A}), the phase diagram parameterized by the bare interactions $J$ and $U_{p\parallel}$ (Eq.~(7)). The renormalization effects are taken into account by solving Eq.~\eqref{eq:rgequations} numerically. We fix the inter-particle distance in the 1d tube to be $50$ nm to be consistent with atomic density at the trap center the experiment.
Here we set the Planck constant  $\hbar$, the atomic mass $M$, and the tunneling $t_{sp}$ as units. 
({\bf B}), shows one example of renormalization flow where $J$ renormalizes from the positive to the negative side, enabling the emergence of the Potts-nematic phase.
({\bf C}), shows one example with $J$ remaining at positive side after renormalization, which then supports a chiral condensate at lattice-rotation symmetric $K$ points. The horizontal axis $\Lambda$ represents the running momentum scale in the renormalization flow. In the numerical results, we also find the renormalization effects in the $p$-orbital channel are quantitatively much stronger than the $s$-orbital channel. 
In our calculation of the bare couplings for  the experimental lattice setup, 
  we find the Josephson coupling $J$ is negligible compared to other interaction strengths, which are comparable to the tunneling $t_{sp}$. 
  This favors the nematic phase after consideration of interaction renormalization. 
}
\label{figs:phasediag}
\end{figure}

\subsection{Field theoretical renormalization effects}
In this supplementary section, we explain why the orbital Josephson coupling is negative despite its tree-level estimate~\cite{1994_Shankar_RMP} is positive for repulsive atoms. We construct the field theoretical action under the standard path integral formalism for the multi-orbital bosonic system as
\bea
\label{eq:action} 
S[\Phi] &=& \int d\tau dz \sum_{{\bf r }{\bf r}'} \Phi^\dag _{\bf r} (z, \tau ) \left[\partial_\tau -\frac{\hbar^2}{2M}\frac{\partial^2}{\partial z ^2} +{\cal H} _{ {\bf r}{\bf r}'}  \right] \Phi_{{\bf r}'} (z, \tau)  \nn \\
&+&  \int d\tau dz\frac{1}{2} U_s \sum_{\bf r \in {\cal B} }\phi_{s, {\bf r}} ^\dag  \phi_{s, {\bf r} } ^\dag \phi_{s, {\bf r}}\phi_{s, {\bf r}}  \\
&+& \int d\tau dz \sum_{ {\bf r }\in {\cal A} }
  \left\{ J  \left[ \phi_{x, {\bf r}} ^\dag \phi_{x,{\bf r}} ^\dag \phi_{y, {\bf r}} \phi_{y, {\bf r}}+ H.c. \right]
+\frac{1}{2} \sum_{\alpha,\beta \in \{x,y\}}U_{p,\alpha \beta} \phi_{\alpha, {\bf r}} ^\dag \phi_{\beta,{\bf r}} ^\dag \phi_{\beta, {\bf r}}  \phi_{\alpha, {\bf r}}
 	\right\}. \nn
\eea
The corresponding partition function reads
$$Z = \int {\cal D} \Phi^* D \Phi  \exp(-S[\Phi^*, \Phi]).$$
Here $\Phi_{\bf r} (z, \tau)$ is  a compact notation for $[ \phi_{x, {\bf r}} (z, \tau) , \phi_{y, {\bf r}} (z, \tau), \phi_{s. {\bf r}} (z, \tau)]^T$, which are fluctuating fields associated with annihilation operators $[\hat{p}_{x, {\bf r}}, \hat{p}_{y,{\bf r}}$, $\hat{s}_{\bf r}]$ in the path integral formalism, and ${\cal H}$ the tunneling matrix.
The associated non-interacting Green function is given by
$G({\bf k}, \omega) = [-i\omega +{\cal H} ({\bf k}) ]^{-1} $,
with
\be
{\cal H} =
\begin{bmatrix}
\frac{\hbar^2k_z ^2 }{2M} -\mu_p  & 0	&\frac{\sqrt{3}}{2} t_{sp} (e^{ik_3} -e^{-ik_2}) \\
0	&\frac{\hbar^2k_z ^2 }{2M} -\mu_p 	& \frac{1}{2} t_{sp}  (e^{ik_3} + e^{-ik_2} -2)\\
\frac{\sqrt{3}}{2} t_{sp} (e^{-ik_3} -e^{ik_2}) & \frac{1}{2}  t_{sp} (e^{-ik_3} + e^{ik_2} -2)	& \frac{\hbar^2k_z ^2 }{2M} -\mu_s
\end{bmatrix} ,
\ee
with $k_3 = {\bf k} \cdot {\bf a}_3 $, and $k_2 = {\bf k} \cdot {\bf a}_2$.
Here $M$ is the atomic mass, and $t_{sp}$ is the tunneling between $s$-orbitals on ${\cal B}$ sites  and the nearby $p$-orbitals on ${\cal A}$ sites.

We have incorporated the continuous degrees of freedom along the tube ($z$-direction) in this field theory. Considering the $C_{3v}$ symmetry, we have
 $U_{p,xx} = U_{p,yy} \equiv U_{p\parallel }$, $U_{p,xy} = U_{p,yx} \equiv U_{p\perp}$, $J = (U_{p\parallel}-U_{p\perp} )/2$.
Introducing the Fourier components of the fields as
$
\phi_{a\in\{x,y,s\}, {\bf r}}(z, \tau) = \int \frac{ d^3{\bf k} d\omega}{(2\pi)^4} \phi_a ({\bf k}, \omega) e^{i(k_x r_x +k_y r_y + k_z z -\omega \tau)} ,
$
the non-interacting Green functions defined by
$G_{ab} ({\bf k}, \omega) = \langle {\phi}^\dag_a({\bf k}, \omega) {\phi} _b ({\bf k}, \omega) \rangle$
are given by the Fourier transform of $\left [\partial_\tau -\frac{\hbar^2}{2M}\frac{\partial^2}{\partial z ^2} +{\cal H}  \right]^{-1} $,
with $\omega$ the Matsubara frequency.

The bare couplings, $U_{p \parallel}$, $U_{p \perp}$, and $J$, in the field theory are determined by matching the interaction energy obtained from the tight binding model in Eq.~\eqref{eq:action} and from the continuous field theory. More specifically, we take a condensate wavefunction for N-particles condensing at a quasi-momentum ${\bf k}=(k_x, k_y)$ of the $n$-th Bloch band, denoted as $|\Psi_{n,{\bf k}}\rangle$. We first calculate its interaction energy $E_{n,{\bf k}}$ according to the continuous theory 
$\frac{2\pi\hbar ^2 a_s} {M} \int d^3 {\bf x} \phi ^\dag ({\bf x}) \phi ^\dag ({\bf x}) \phi ({\bf x}) \phi({\bf x})$, 
with $a_s$ the scattering length, $\phi({\bf x})$ the field operator describing the continuous degrees freedom of atoms. 
The obtained interaction energy is dependent on the scattering length. We then calculate the interaction energy according to the tight binding model in Eq.~\eqref{eq:action}, $E_{n, {\bf k}} ^{\rm TB}$, as a function of the bare couplings $U_{p,\parallel}$, $U_{p, \perp}$, and $J$. These couplings are then obtained by a least square fit that minimizes 
a cost function $F = \sum_{n} \int \frac{ d^2 {\bf k} } {(2\pi)^2 }  |E_{n, {\bf k}} - E_{n, {\bf k}} ^{\rm TB}|^2$. In our calculation, we find $\sqrt{F}$ is below $5\%$ in units of $t_{sp}$. 
We note here that this approach would reduce to calculating the Wannier function overlap for a single-band tight-binding model~\cite{1998_Zoller_PRL}.

To proceed we introduce two functions for compactness,
\bea
&h (k_x, k_y)  = \sqrt{3-\cos k_1 - \cos k_2 -\cos k_3 }  &,\nn \\
&
\Delta (k_x, k_y) = \sqrt{[\mu_s-\mu_p]^2 /4+ [ h(k_x, k_y)t_{sp} ]^2 } &.  \nn
\eea
The Green function is diagonalized to be
$G({\bf k}, \omega) = {\cal U} {\cal U}' D({\bf k}, \omega) {\cal U}'^{\dag} {\cal U} ^\dag  $, with
\bea
& {\cal U} =
\begin{bmatrix}
-\frac{(e^{-ik_3} + e^{ik_2} -2)/2}{ h(k_x,k_y) } 	
			& \frac{\sqrt{3} (e^{ik_3} -e^{-ik_2})/2}{ h(k_x,k_y)}	&0\\
\frac{\sqrt{3} (e^{-ik_3} -e^{ik_2})/2}{ h(k_x,k_y) }
		&\frac{(e^{ik_3} + e^{-ik_2} -2)/2}{ h(k_x,k_y)} &0 \\
0 	&0	& 1
\end{bmatrix}
&  \\
&
{\cal U}' =
\begin{bmatrix}
1	&0	&0 \\
0	&\cos (\vartheta/2)	& -\sin(\vartheta/2) \\
0	&\sin(\vartheta/2)	&\cos(\vartheta/2)
\end{bmatrix}
& \nn \\
&
D({\bf k}, \omega) = {\rm diag}
\left[ f_0 ({\bf k}, \omega),  f_+  ({\bf k}, \omega),  f_-({\bf k}, \omega)\right]
&.
\eea
Here
$\vartheta$ is defined by
$\cos(\vartheta) = \frac{(\mu_s- \mu_p)/2}{\Delta (k_x, k_y) }$,
and $\sin(\vartheta) =\frac{ h(k_x, k_y)t_{sp} }{\Delta (k_x, k_y) }$, and
$ f_{\gamma =0, +, -}  = [-i\omega + \epsilon_\gamma ({\bf k})]$,
with
$ \epsilon_0 = \frac{\hbar^2 k_z^2 }{2M} - \mu_p$, and
$
\epsilon_\pm  = 	
 \frac{\hbar^2 k_z^2}{2M}  -\frac{ \mu_s + \mu_p}{2} \pm  \Delta(k_x, k_y).
$
The Green functions $G_{ab} ({\bf k} ,\omega) \equiv \langle \phi_a ^\dag (\bf k, \omega) \phi_b ({\bf k}, \omega) \rangle$,  are obtained as
\bea
G_{xx} ({\bf k}, \omega) &=&
	\sin^2 (\varphi) f_0  + \cos ^2 (\varphi)  \left[ \cos^2 (\vartheta/2) f_+ + \sin^2 (\vartheta/2) f_- \right], \nn \\
G_{yy} ({\bf k}, \omega) &=& 
	\cos^2 (\varphi) f_0  + \sin ^2 (\varphi)  \left[ \cos^2 (\vartheta/2) f_+ + \sin^2 (\vartheta/2) f_- \right], \nn \\
G_{xy} ({\bf k}, \omega) 
&=& \frac{\sqrt{3}(e^{-ik_3} + e^{ik_2} -2)(e^{ik_3} -e^{-ik_2})}{4[h (k_x, k_y)]^2}
\left[ \cos^2 (\vartheta/2) f_+ + \sin^2 (\vartheta/2) f_- -f_0 \right]  \nn \\
G_{ss} ({\bf k}, \omega) &=& 
\cos^2 (\vartheta/2) f_- + \sin^2 (\vartheta/2) f_+  , \nn
\eea
with $\varphi$ defined by $\cos^2 (\varphi) = \frac{3|e^{ik_3} -e^{-ik_2} |^2}{4[h(k_x,k_y)]^2 } $.
The orbital mixing between $p_x$ and $p_y$, $G_{xy}$, is mediated by the $s$-orbital, which vanishes at the limit of $t_{sp}\to 0$.

Introducing a running energy scale $\Lambda$ which is continuously  decreased from an initial $\Lambda_0$,
the couplings in a renormalized mean field theory can be derived by continuously integrating out the high energy modes with momentum $|k_z| \in [\Lambda -\Delta \Lambda , \Lambda]$~\cite{2011_Kivelson_PRB}.
The renormalization of couplings among the low-energy modes is determined according to
$e^{-S_{\rm eff} [\Phi^{*}_<, \Phi_{<}] } = \int D\Phi^{*}_> D\Phi_> e^{-S[\Phi^*, \Phi]}$, with $\Phi_>$ ($\Phi_<$) referring to high (low) energy modes. Keeping one-loop Feynman diagrams, the renormalization is obtained in terms of kernel integrals
$$
I_{abcd} (\Lambda) =\frac{1}{\pi} \int \frac{dk_x dk_y d\omega}{(2\pi)^3}G_{ab}(k_x, k_y, \Lambda, \omega) G_{cd} (-k_x, -k_y, -\Lambda, -\omega),
$$
as
\bea
 \frac{\Delta U_s } {\Delta \Lambda} &=& -[U_s (\Lambda)] ^2 I_{ssss}  (\Lambda) \\
 \frac{\Delta [U_{p\parallel} + 2J] }{\Delta \Lambda} &=& - [U_{p\parallel} (\Lambda) + 2J (\Lambda)] ^2 [I_{xxxx}  (\Lambda)+ I_{xyxy} (\Lambda)]  \\
 \frac{\Delta U _{p\perp} }{\Delta \Lambda} &=& -[ U _{p\perp} (\Lambda) ]^2 [I_{xxxx} (\Lambda) - I_{xyxy} (\Lambda)],
\eea

Here we discuss several key properties of these integrals.
Due to time-reversal symmetry, we have $ G ({\bf k}, \omega) = G^*(-{\bf k}, -\omega)$.
It follows that these terms
\be
I_{abab} (\Lambda) >0.
\ee
Considering $p$-orbitals form a two-dimensional representation ($E$) of  the $C_{3v}$ rotation group, we have
\be
I_{xxxx} = I_{yyyy}, \,\,\,\,\,
I_{xyyx} =  I_{xxxx} - I_{xxyy} - I_{xyxy}.
\ee
We have used these symmetries in the derivation which help simplify the form of the  renormalization equations.
 It is worth emphasizing here that the term $ I_{xyxy} $ comes from the orbital mixing  between $p_x$ and $p_y$ mediated by the $s$ orbital, which vanishes at the limit of $t_{sp} \to 0$. This orbital mixing makes the $p$-orbital bosons in the hexagonal lattice drastically distinctive from that in the square lattice.

The integral over the frequency $\omega$ can be carried out analytically using
$
\int \frac{d\omega}{2\pi}  f_\gamma({\bf k}, \omega) f_{\gamma'}(-{\bf k}, -\omega) = \frac{\Theta (\epsilon_{\gamma'} (-{\bf k})) - \Theta(\epsilon_{\gamma} ({\bf k})) }{\epsilon_\gamma ({\bf k}) + \epsilon_{\gamma'} (-{\bf k})}
$
with $\Theta$ the heavyside step function.
Then we know that the leading order term in the kernel integrals $I_{abcd}$ scale as $1/\Lambda^2$ for high-energy modes.
Introducing a running scale $l$ by $\Lambda = \Lambda_0 l^{-1}$, the leading renormalization of the low-energy couplings is described by a flow equation
\bea
 \label{eq:rgequations}
  \frac{d U_s } {d l} &=& -\Lambda_0 [U_s (\Lambda)] ^2 [I_{ssss}  (\Lambda_0 )] \nn \\
 \frac{d [U_{p\parallel} + 2J] }{dl } &=& - \Lambda_0[U_{p\parallel} (\Lambda) + 2J (\Lambda)] ^2 [ I_{xxxx}  (\Lambda_0)+ I_{xyxy} (\Lambda_0)]  \\
 \frac{d U _{p\perp} }{dl } &=& -\Lambda_0 [ U _{p\perp} (\Lambda) ]^2 [I_{xxxx} (\Lambda_0) - I_{xyxy} (\Lambda_0)] . \nn
\eea
Then we get  an invariant in the renormalization,
\be
C_{\rm Renorm} = \frac{I_{xxxx}+I_{xyxy}}{U_{p,\parallel} (\Lambda)  - 2J (\Lambda)}  -\frac{I_{xxxx}-I_{xyxy}}{ U_{p,\parallel} (\Lambda)  + 2J(\Lambda) } .
\ee
With a bare positive coupling $J(\Lambda_0)>0$, we have $C_{\rm Renorm} >0$,  and the running couplings  would always renormalize to  a point of
$
U_{p\parallel} = U_{p \perp} = \frac{2I_{xyxy}}{C_{\rm Renorm}}  , J = 0,
$
and then flow to the negative side of $J$.
We thus establish the tendency of the Josephson coupling renormalizing to a negative value with the field theory analysis. This property is generic provided that $I_{xyxy}$ (or equivalently the off-diagonal Green's function $G_{xy}$) is finite, or in physical words the orbital mixing between $p_x$ and $p_y$ is finite. The characteristic renormalization flow is shown in
\textit{Fig. 4 in the main text}.
The renormalization theory explains why the orbital Josephson coupling $J$ is negative in the renormalized mean field theory, as required  by the Potts-nematicity.

{
We also carry out numerical calculation of the renormalization flow and obtain a phase diagram according to a renormalized mean field theory~\cite{2011_Kivelson_PRB}. The results are shown in Fig.~\ref{figs:phasediag}. We find the Potts-nematic phase indeed occurs even in the region with a positive Josephson coupling. It is also worth noting here that the renormalization effects of the interaction in the $p$-orbital channel are more pronounced than that in the $s$-orbital channel.
}

\end{widetext}

\end{document}